# Topology and Nematic Ordering I: A Gauge Theory

Paul E. Lammert
*Ecole Supérieure de Physique et de Chimie Industrielles,*
*10, rue Vauquelin, 75231 Paris Cedex 05, France*

Daniel S. Rokhsar
*Department of Physics, University of California, Berkeley, CA 94720*

John Toner
*IBM T. J. Watson Research Center, Yorktown Heights, NY 10598*

We consider the weakly first order phase transition between the isotropic and ordered phases of nematics in terms of the behavior of topological line defects. Analytical and Monte Carlo results are presented for a new coarse-grained lattice theory of nematics which incorporates nematic inversion symmetry as a local gauge invariance. The nematic-isotropic transition becomes more weakly first order as disclination core energy is increased, eventually splitting into two continuous transitions involving the unbinding and condensation of defects, respectively. These transitions are shown to be in the Ising and Heisenberg universality classes. A novel isotropic phase with topological order occurs between them.

## I. INTRODUCTION

The modern theory of critical phenomena [1] emphasizes the role of symmetry and dimensionality in determining the long-wavelength behavior at phase transitions. From this point of view, the properties of the ferromagnet-to-paramagnet [1] and nematic-to-isotropic [2] order-disorder phase transitions might be expected to be quite similar. In both cases, a continuous rotational symmetry is broken, leading to the same sort of long-wavelength Goldstone modes, spin waves and director modes, respectively. Because of this similarity, an expansion [3] in $\epsilon = d - 2$ for spatial dimensions $d$ near two predicts that both transitions are continuous and have identical exponents to all orders in $\epsilon$.

Prior to our earlier work [4], it was generally believed that this prediction was wrong in three dimensions, where the ferromagnet to paramagnet transition is continuous, whereas all observed nematic to isotropic transitions are first order, albeit usually only weakly so.

These empirical facts are in accord with the predictions of Landau theory [2], although that treatment fails to explain the ubiquitous weakness of the first order nematic transition. The Landau approach treats these two order-disorder transitions using qualitatively different order parameters, and apparently disregards the inherent similarities of the ordered states. In this paper we will focus on the role of topological defects in the nematic phase transition, since "disclinations" – the line defects which occur in nematics but are not allowed in magnetic systems – are the sole topological distinction between the two kinds of ordering.

This paper presents a thorough analysis of a new model [4] of the three-dimensional nematic-isotropic ($N/I$) transition in which a pivotal role is played by disclinations. Our model allows us to independently vary the local nematic stiffness and the disclination core energy. When the defect core energy is large, our nematic contains only a few, thermally activated, small disclination loops, and is therefore similar to the magnet, where such loops are topologically forbidden.

We find that sufficiently strong suppression of defects leads to behavior that contradicts the predictions of Landau theory. In particular, the first order nematic-isotropic transition splits into two *continuous* transitions, with a novel intervening "topologically ordered" isotropic phase (see figure 1). The transition from the conventional isotropic phase (denoted by '$I$') into the new topologically ordered isotropic phase ('$T$') belongs to the universality class of the three-dimensional Ising model, and that from the topologically ordered phase to the nematically ordered phase ('$N$' for "nematic") to the three-dimensional, three-component ($n = 3$) Heisenberg universality class. Recent experiments by Roux *et al.* [5] may have revealed such a phase diagram.

The organization of this paper is as follows. In section II we further discuss the the conflicting approaches to the $N/I$ transition based on Landau theory and $2 + \epsilon$ expansion, and describe the disclinations which are the sole topological distinction between ferromagnets and nematics. Motivated by these considerations, we present in section III two models which explicitly suppress these defects, with the purpose of making the nematic more like the ferromagnet. The phase diagram of this second model (figure 3), or more specifically the existence of three distinct phases and the natures of the transitions between them, is determined in sections IV-VII. Analysis of three limits which reduce to already understood models is presented in section IV, followed, in section V by a characterization of the three phases in terms of *defect line tension*, and the introduction of order parameters in section VI. The robustness of the exact limits analyzed in section IV is demonstrated in section VII, thereby establishing in particular that the nematic to physically isotropic (but topologically ordered) transition remains continuous and in the Heisenberg universality class over a non-vanishing range of parameters. This section is fairly dense and can be skipped on first reading. Section VIII shows that perturbations such as space-spin coupling that are absent from our model yet present in real nematics do not alter our conclusions. Our Monte Carlo results on the new nematic gauge theory, including a finite size scaling analysis of the phase transitions, are presented in section IX, and corroborate the analysis of earlier sections. A separate paper [6], Paper II, presents detailed predictions for critical phenomena at the new transitions predicted here.

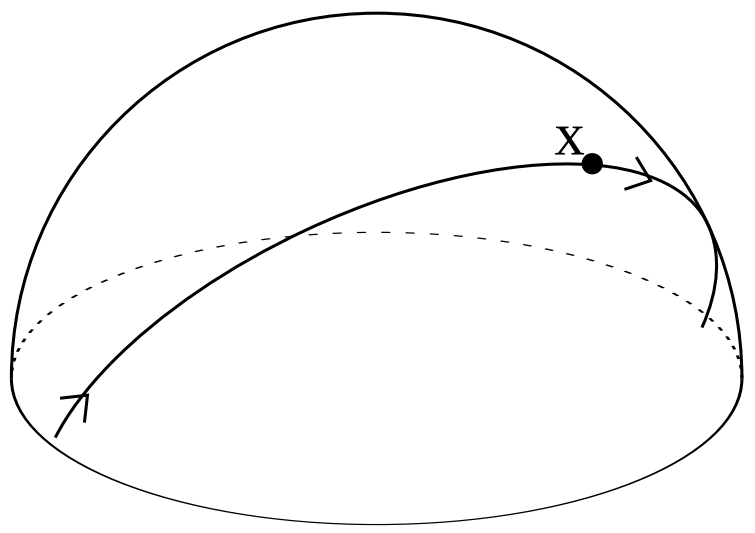

FIG. 1. The projective plane $\mathbb{RP}^2$. Points on the equator and in the 'missing' hemisphere are identified with their antipodes, so that it does not have a boundary, contrary to casual appearances. A *closed*, but non-contractible, path starting and ending at point $x$ is depicted.

## II. SYMMETRY, EPSILON EXPANSION, AND LANDAU THEORY

**Fluctuations.** Ferromagnets and nematics both spontaneously break a continuous rotation symmetry: the elementary moments of a ferromagnet are preferentially aligned parallel to a common *vector* in "spin space," while in a nematic the long axes of the constituent molecules preferentially align with a common *axis* in physical space. By "axis" we mean a headless vector, usually called a "director." Note that there are two directions associated with an axis, so the nematic has a local "up-down" symmetry which is lacking in the ferromagnet.

The "order parameter space" of an ordered medium contains the set of "directions" that are associated with the broken symmetry. The order parameter space for the Heisenberg magnet is the unit sphere $S^2$. For the nematic, the order parameter space is the projective plane $\mathbb{RP}^2$, which is simply the unit sphere with antipodal points identified. There is evidently a two-to-one mapping from $S^2$ to $\mathbb{RP}^2$ which is a local isometry. Thus small fluctuations of both order parameters therefore have similar phase space measures.

In an ordered phase, the effective free energy for small fluctuations can be written in terms of gradients of the order parameter. Since the nematic and ferromagnetic order parameter spaces are locally isometric, the associated long-wavelength physical properties are also the same. Indeed, the Frank free energy density for nematics,

$$\mathcal{F}_{\rm grad}(\mathbf{r}) = \frac{1}{2}[K_1(\nabla\cdot\mathbf{n})^2 + K_2(\mathbf{n}\cdot(\nabla\times\mathbf{n}))^2 + K_3(\mathbf{n}\times(\nabla\times\mathbf{n}))^2], \qquad (1)$$

is an acceptable spin-wave Hamiltonian for a ferromagnet with space-spin coupling. (If all of the $K_i$ are equal, eq. (1) becomes simply $(K/2)(\partial_\alpha n_\beta)(\partial^\alpha n^\beta)$, which is invariant under independent rotations of space and of the order parameter.)

An immediate corollary of this discussion is that any treatment of nematic ordering which considers only nonsingular fluctuations about a uniform, nematically ordered state will necessarily be identical to the corresponding treatment of a ferromagnet. This is precisely the approach of the $\epsilon$ expansion [7]. The existence of a critical fixed point of order $\epsilon$ implies that both transitions can [8] be continuous, and would share the same universality class.

**Topology.** Although the local fluctuations of the ordered nematic and ferromagnetic states are quite similar, certain configurations of the two systems are very different. Suppose for example that we arbitrarily assign an arrowhead to the director at some point in a nematic, and try to extend this smoothly to a continuous vector field that is consistent with the given director field. For small fluctuations this is clearly possible; traversing a closed path in physical space maps out a closed path on the unit sphere $S^2$. Such configurations are "homotopically trivial." Note that all smooth magnetic configurations are trivial in this sense. Homotopically trivial nematic configurations can therefore be placed in correspondence with related magnetic configurations. Since their energies will be both be governed by the elastic energies of eq. (1), we expect these configurations to make similar contributions to the partition function of their respective systems.

The nematic configurations which cannot be related to magnetic configurations in this manner are those for which an ambiguity in sign arises at the completion of a closed path in physical space. That is, the image in $\mathbb{RP}^2$ of a closed path in the physical system will be closed if the director field is continuous, but its "lift" to $S^2$ travels from one point to its antipode and is not closed. A closed curve in the nematic that corresponds to a homotopically nontrivial loop in $\mathbb{RP}^2$ necessarily encircles a singularity of the order parameter, since shrinking the path in real space induces a deformation of the corresponding path in the order parameter space. Such a configuration is said to contain a topological line defect known as a "disclination." The topological character of these defects endows them with considerable dynamical stability. This essential distinction between nematic and magnet is quantified by the fundamental group $\pi_1$. The fundamental group $\pi_1(S^2)$ of the sphere is the trivial group with only an identity element, whereas that of the projective plane, $\pi_1(\mathbb{RP}^2)$, is the group $\mathbb{Z}_2$ (integers mod 2 under addition).

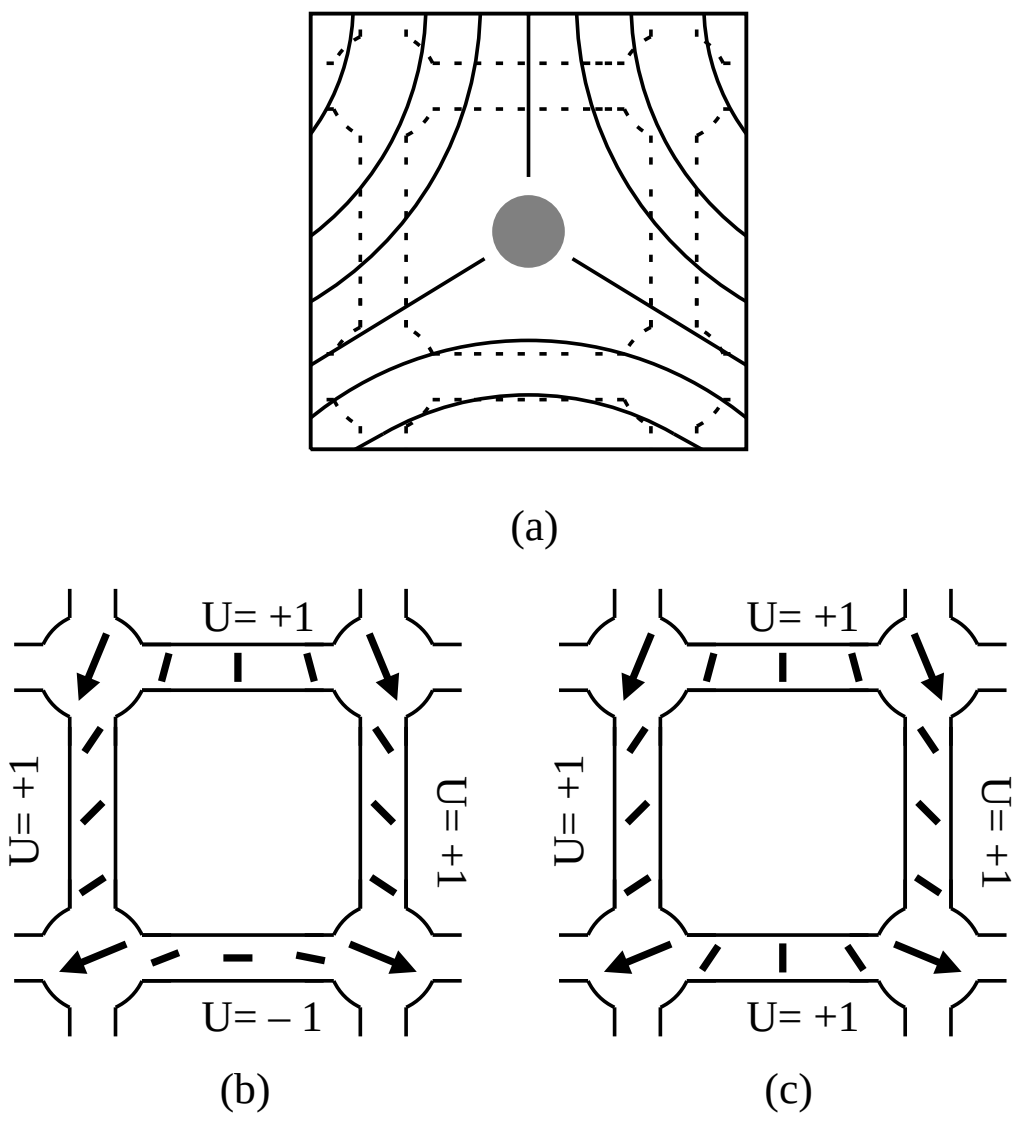

FIG. 2. (a) A cross section through a disclination. When the defect is encircled, the local molecular axis rotates through $180^o$. On a lattice the defect can occur inside a plaquette. The presence (b) or absence (c) of a defect depends on the product of the $U_{ij}$ around the plaquette.

A cross-section of a nematic configuration with a disclination is depicted in figure 2. Related configurations in the magnet have very large energies, because there are 180 degree discontinuities in the corresponding vector field. Thus configurations containing defects will make distinct contributions to the partition functions of the nematic and magnet, and provide a possible explanation for the fundamental difference between the two order-disorder transitions. These disclinations are very important for the nature of the nematic to isotropic transition. That this is so is suggested by earlier work (for a very readable review see reference [9]; also see references [10–12]) showing that many phase transitions proceed by an unbinding of topological defects.

The most famous example is the explanation of the transition in the two-dimensional XY model as an unbinding of point vortices by Kosterlitz and Thouless. [13,14] In three dimensions, both the XY model and the superconductor can be cast as problems of interacting vortex loops [15]. This produced the discovery that the superconducting-normal transition can be continuous (as experimentalists are well aware!), in contrast to the prediction of the $4 - \epsilon$ expansion that fluctuations of the electromagnetic field drive a first order [16] transition. Monte Carlo studies [17] have found that suppression of these loops (by adding a sufficiently large core energy) constrains the model to remain in its ordered phase, and prevents a phase transition to a disordered state.

In the three-dimensional Heisenberg model, there are no stable line defects, but point defects ("hedgehogs") are possible. Suppression of these point defects in the three-dimensional Heisenberg model has been studied in simulations by Lau and Dasgupta [18], in which the transition was destroyed by a sufficiently large energetic penalty for hedgehogs. Fucito and Solomon [19] likewise found that the three dimensional XY transition was destroyed by suppressing vortex lines.

**Landau theory.** In Landau theory, the free energy is expanded in powers of an "order parameter [21]." Since the nematic has global inversion symmetry its order parameter cannot be a vector: there is no macroscopic direction selected by the ordered state. There is, however, a preferred axis, which leads to anisotropy in *e.g.* magnetic and dielectric susceptibility. The order parameter can then be chosen to be a traceless symmetric tensor $Q_{\alpha\beta}$. (For this discussion of Landau theory we restrict ourselves to three-dimensional systems, so $\alpha$ and $\beta$ here run from 1 to 3.) . For weak nematic order $Q_{\alpha\beta}$ is proportional to the anisotropy of the susceptibilities and light scattering, and can be thought of as the quadrupole moment of the local distribution of molecular orientations.

The Landau theory of the nematic-isotropic transition is readily constructed in terms of $Q$ by writing down all rotationally invariant scalars. To fourth order, the Landau free energy is

$$\mathcal{F}_{\rm LG} = a(T - T_0)\,{\rm Tr}\,[{\rm Q}^2] - {\rm b}\,{\rm Tr}\,[{\rm Q}^3] + 2{\rm c}\,{\rm Tr}\,[{\rm Q}^4]. \quad (2)$$

(Note that the fourth order term is unique, since $({\rm Tr}\ {\rm Q}^2)^2 = 2\,{\rm Tr}\ {\rm Q}^4$ for any traceless symmetric three-by-three matrix.) $Q$ has five independent components, and the quadratic term is simply the sum of their squares. Thus, if $b = 0$, a system governed by this free energy will undergo a second-order transition at $T_0$ in the $O(5)$ universality class. The cubic term is relevant at the associated fixed point, however, and there certainly appears to be no reason for $b$ to vanish generically. When $b \neq 0$ the system undergoes a first order transition at

$$T_* = T_0 + \frac{b^2}{24ac}. \quad (3)$$

For a magnet, however, the corresponding free energy cannot have a cubic term, since inverting the magnetization is a symmetry of the system and therefore cannot change the free energy.

For $b \neq 0$, the free energy (2) is minimized by a *uniaxial* tensor

$$Q_{\alpha\beta} = S(n_\alpha n_\beta - 1/3\delta_{\alpha\beta}), \quad (4)$$

where $\mathbf{n}$ is a unit director. The free energy then reduces to

$$\mathcal{F}_{\mathrm{LG}} = \frac{2}{3}a(T - T_o)S^2 - \frac{2}{9}b\,S^3 + \frac{4}{9}c\,S^4. \tag{5}$$

Uniaxiality is stable against the introduction of higher order terms in eqn. (2) for sufficiently weak nematic order.

Nematic transitions observed in nature are first order (score one for Landau), but generally only *weakly* so [2]. This is revealed by light scattering, which shows nearly-critical fluctuations, and means that the transition temperature $T^*$ is close to the spinodal $T_0$. The jump in the order parameter at the transition,

$$S(T^*_-) = \frac{b}{4c}, \tag{6}$$

is *not* necessarily small. The Landau theory gives no clue as to why the transition should generally be so weak despite significant variation in microscopic properties, though it has been suggested [2] that the aspect ratio of the constituent molecules may provide a natural "small parameter."

**Motivation for our work.** We have seen that the $\epsilon$ expansion, which ignores disclinations, predicts a continuous nematic-to-isotropic transition, while the Landau theory, which implicitly includes disclinations (since it averages, albeit crudely, over all configurations of the full nematic order parameter, including those with defects), predicts a first-order transition. Is it, then, the disclinations that drive the nematic-to-isotropic transition first-order? And if so, can the continuity of the transition be restored by suppressing these defects? Our answer to both questions is a resounding "yes!" These conclusions are reached by considering a new theory of nematics which incorporates both spin-wave and topological fluctuations on an equal footing. We find that the strength of the first-order nematic-to-isotropic transition is reduced as the defect core energy is increased. For large enough defect suppression the transition splits into a pair of continuous transitions, with a qualitatively new intervening phase (see figure 3). This new phase is isotropic ($\langle Q_{\alpha\beta}\rangle = 0$), like the fully disordered phase, but possesses distinct topological characteristics.

**Experimental verification.** To test these ideas experimentally one must have *independent* control over the disclination core energy, corresponding to $K$, and the local nematic interaction strength corresponding to $J$. This is likely to be difficult, since in real materials both microscopic energies arise from the same interactions and entropic configurations. Nevertheless one can imagine a "defectophilic" impurity which would accumulate in defect cores, and thus increase their entropic cost. Similarly, long "defectophobic" molecules might align with the nematogens and make formation of defects more difficult. Observation of weakening (strengthening) of the first order transition as defects are suppressed (favored) would provide partial support for our scenario.

Another approach is to add objects to the nematic which disorder it *without* favoring the creation of disclinations. In this way, the nematic order is destroyed at a lower temperature, where disclinations are effectively suppressed. (That is, $E_c/k_BT$ increases while the defect core energy $E_c$ remains the same.) This effect may explain the experiments of Roux *et al.* [5], who added small polystyrene spheres to a nematic. These spheres favor the creation of point defects (hedgehogs), and therefore tend to destroy nematic order. Roux *et al.* appear to observe critical opalescence, suggesting a continuous transition between two isotropic phases for this system; it is tempting to identify this transition with the $I/T$ transition in figure 3. Detailed experimental tests of the critical properties described in Paper II would confirm this identification.

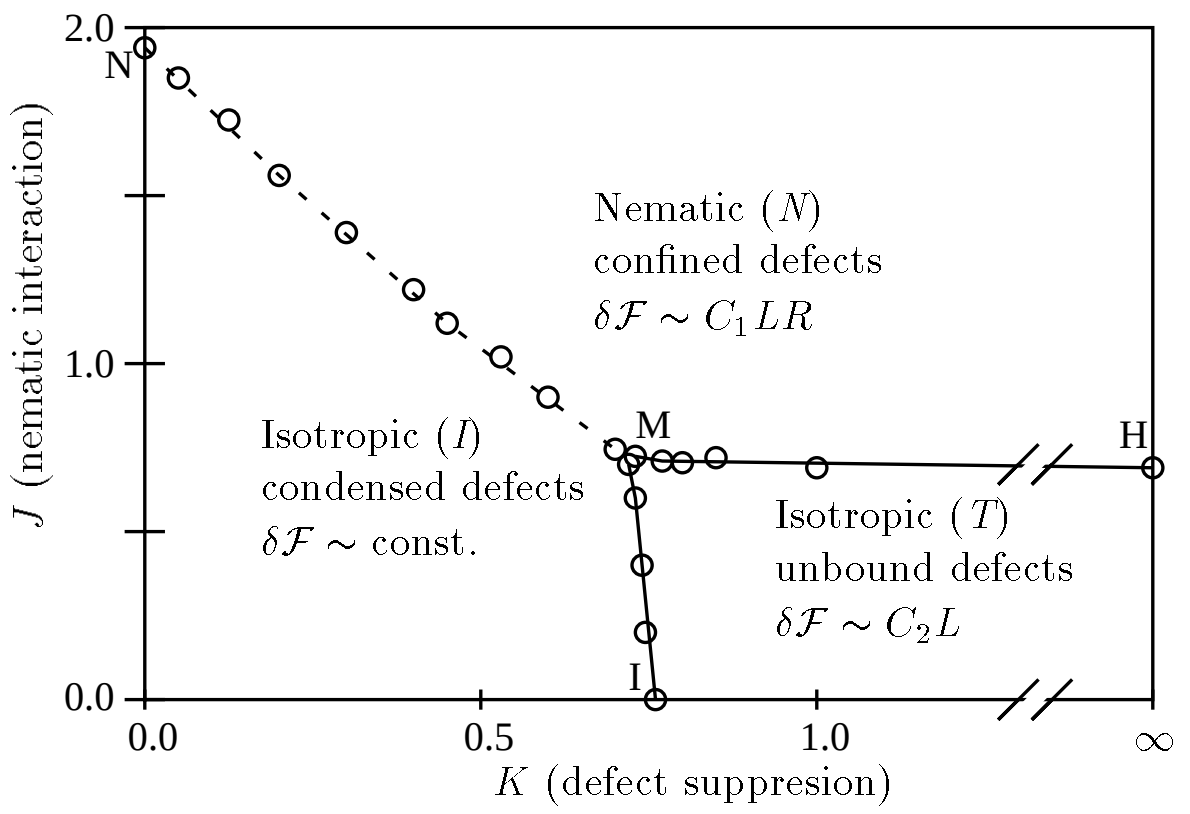

FIG. 3. Phase diagram for the $\mathbb{Z}_2$ gauge plus three-component spin lattice model. The phase boundaries are very straight except near their intersection. The two nematically disordered phases are not distinguishable by any local quantity. The dotted line denotes a first-order transition and the unbroken lines continuous transitions.

Unfortunately, the new topologically ordered phase we predict will be difficult to observe directly because it is isotropic, and therefore quite similar to the conventional isotropic state. Nevertheless, the passage into and out of this new phase can be recognized by the associated critical singularities. These singularities are determined by the universality classes of the transitions, which are three-dimensional Ising for the transition between the fully disordered and topologically ordered phases ($I/T$) and three-dimensional Heisenberg for the transition between topologically and nematically ordered states ($T/N$). Detailed predictions can be found in Paper II.

## III. LATTICE MODELS FOR NEMATIC SYSTEMS

In this section we introduce two lattice models for nematic media which explicitly include energetic penalties for disclinations: a "spin-only" model and a lattice gauge theory. The behavior of the spin-only model is less interesting, but it clarifies the motivation of for the lattice gauge theory. Monte-Carlo studies of this latter model is reported in section IX. Analytic results for the lattice gauge theory will be studied in more detail in sections IV and VII.

### A. Modified Lebwohl-Lasher Model

The simplest lattice model of the nematic-isotropic transition was introduced by Lebwohl and Lasher [22], and is based on the continuum Maier-Saupe model [2]. The local nematic degrees of freedom are represented by unit *vectors* $\mathbf{S}_i$ that are assigned to the sites of a regular lattice (cubic for convenience). The free energy of a configuration of such spins is given by

$$\beta\mathcal{H}_N = -J_N \sum_{\langle i,j\rangle} (\mathbf{S}_i \cdot \mathbf{S}_j)^2 \tag{7}$$

where $\langle i,j\rangle$ denotes nearest-neighbor pairs of lattice sites. Since the spins occur quadratically, there is *local* inversion symmetry, as appropriate for a nematic model. That is, the spin at site $i$ may be negated without changing the energy of the system. This local symmetry foreshadows the introduction of a gauge description of nematics (see below).

To consider topological defects, we introduce a defect counting operator

$$\mathcal{D}_{ijkl} \equiv \frac{1}{2}\left\{1 - \mathrm{sgn}[(\mathbf{S}_i \cdot \mathbf{S}_j)(\mathbf{S}_j \cdot \mathbf{S}_k)(\mathbf{S}_k \cdot \mathbf{S}_l)(\mathbf{S}_l \cdot \mathbf{S}_i)]\right\}, \tag{8}$$

which is unity if a defect pierces the plaquette $(ijkl)$, and zero otherwise. This four-spin operator evidently retains the local inversion symmetry of eqn. (7).

If the term in square brackets vanishes, then by a process of flipping spins (to which the *nematic* configuration, hence the energy, is insensitive), we can make all the factors $(\mathbf{S}_i \cdot \mathbf{S}_j)$ positive. Then the smooth interpolation which takes the shortest route between the corresponding values for the director (on $\mathbb{RP}^2$) is homotopically trivial. Thus, there is no disclination threading the plaquette $(ijkl)$. In contrast, in the presence of a defect there is always one leftover negative factor, and the defect number is unity. The associated path on $\mathbb{RP}^2$ is nontrivial.

A defect core energy is then

$$\beta\mathcal{H}_D \equiv K_D \sum_{\Box} \mathcal{D}_{ijkl} \tag{9}$$

where the sum extends over all elementary plaquettes $\Box = (ijkl)$. A Monte Carlo study shows that the transition weakens (as measured by the order parameter discontinuity) as $K_D$ increases. There are only two states (nematic and isotropic) in the phase diagram of the model of eqns. (7-9), and the transition between them is always first order, as expected by Landau theory. For large enough $K_D$, however, the system remains ordered even for vanishing bare nematic stiffness $J_N$. [23] Thus the elimination of defects leads to a long-wavelength renormalized nematic stiffness which is larger than the value required for long-range order to appear. This result mimics related work on the three-dimensional XY [17] and Heisenberg [18] models, where elimination of defects leads to long-range order even in the absence of a bare order-parameter stiffness. This model, therefore, nowhere exhibits a *continuous* nematic-to-isotropic transition. In the next subsection, we consider a different model, but in the same spirit as this one, which does achieve this goal.

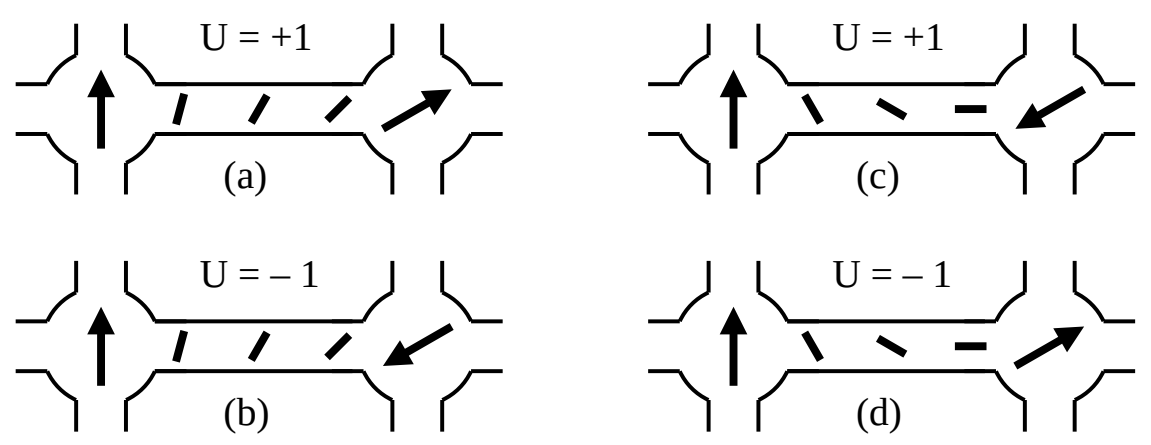

FIG. 4. The link variables $U_{ij}$ provide information on the rotation of the local molecular axes between sites $i$ and $j$.

### B. Lattice Gauge Theory

The model of equations (7-9) lives on a lattice. To deal with topological defects, however, we had to consider what happens *between* lattice sites. To take this into account from the beginning, we envision the lattice as embedded in a (continuum) nematic fluid. A coarse-grained director at each lattice point can be constructed from the mean molecular axis within a region whose radius is somewhat less than the lattice spacing. This procedure can be carried out as long as the correlation length is larger than the lattice spacing. In our model this local molecular axis is associated with a *vector* $\mathbf{S}_i$ at each site $i$, as in the Lebwohl-Lasher model. Once again, the sign of $\mathbf{S}_i$ at each site is arbitrary.

Now consider the variation of the local molecular axes *between* two lattice sites $i$ and $j$ (see figure 4). Beginning at $\mathbf{S}_i$ on the unit sphere and following the variation of the director, we trace out a curve which ends at either $\mathbf{S}_j$ or $-\mathbf{S}_j$. Thus we arrive at two homotopy classes of paths as discussed in the introduction. Those in each class are deformable into one another (while keeping endpoints fixed). In our lattice description of the nematic fluid, we must retain this information regarding the director field between sites; it is needed to define topological defects, which require a notion of continuity. The presence or ab-sence of a defect within a plaquette is determined by the homotopy class of the director field around it, which in turn is determined by splicing together the classes corresponding to the links making up the plaquette.

The local degree of freedom associated with each link $(ij)$ is represented by $U_{ij}$, which can be either $+1$ or $-1$. A value $+1$ of $U_{ij}$ means that one can assign a continuously varying *orientation* to the molecules between $i$ and $j$ which matches up with both $\mathbf{S}_i$ and $\mathbf{S}_j$ (fig. 4a). If the orientation can be chosen to match either $\mathbf{S}_i$ or $\mathbf{S}_j$, but not both, $U_{ij}$ is $-1$ (fig. 4b).

This approach differs crucially from the modified Lebwohl-Lasher model of eqns. (7-8), which implicitly assumes that the local nematic axis follows the shortest route between its values on the lattice sites. If the lattice is not too coarse, this is justifiable by energetic considerations which suggest that these configurations will dominate the partition function. When defects are strongly suppressed, however, this need no longer be the case. The extra energy cost incurred by "taking the long way" may be more than compensated by avoiding the imposed defect core energy. Thus in a strongly defect-suppressed regime, we must keep track of configurations in both the defective and defect-free classes.

By introducing the link variables $U_{ij}$, the local $\mathbb{Z}_2$ invariance of a nematic fluid is expressed as a *local gauge invariance.* Choosing $\phi_i = \pm 1$ independently for each site $i$, the energy of any configuration is unaltered if the simultaneous transformations

$$\begin{aligned} \mathbf{S}_i &\to \phi_i \mathbf{S}_i \\ U_{ij} &\to \phi_i \phi_j U_{ij} \end{aligned} \tag{10}$$

are performed on the site and link variables. The simplest Hamiltonian including a defect-suppression term which respects this local gauge symmetry is

$$\beta\mathcal{H} = -J \sum_{\langle ij \rangle} U_{ij} \mathbf{S}_i \cdot \mathbf{S}_j - K \sum_{\Box} U_{ij} U_{jk} U_{kl} U_{li}, \tag{11}$$

where the second sum is over all elementary plaquettes $\Box = (ijkl)$. The partition function is found from $e^{-\beta\mathcal{H}}$ by integrating over spins $\{\mathbf{S}_i\}$ and summing over link variables $\{U_{ij}\}$.

The first term in eqn. (11) is a nematic interaction which favors minimal variation of the *director* along link $(ij)$. For example, the configurations depicted in figs. 3a and 3c have lower energy than those in 3b and 3d.

The second term is a defect core energy analogous to eqn. (8). If the product of the link variables $U_{ij}$ around a plaquette is $+1$, then there is a smooth pattern of *oriented* molecular axes along the links which does not use the head-to-tail symmetry of the nematic. This pattern can be continuously extended to the entire interior of the plaquette, indicating the absence of a disclination. If the product of the link variables $U_{ij}$ around a plaquette is $-1$, however, the local molecular axis rotates by $180^o$ as the plaquette is encircled (compare figures 2b and 2c). This is most easily seen by using a gauge transformation to set as many of the links on the plaquette to $+1$ as possible (all of them if the product of links is $+1$, all but one otherwise).

How do the defect suppression terms in equations (8) and (11) compare? From the first term in eqn. (11), we see that it is energetically preferable for $\mathrm{sgn}(\mathbf{S}_i \cdot \mathbf{S}_j)$ to be equal to $U_{ij}$. Thus $U_{ij}U_{jk}U_{kl}U_{li}$ (sites $i, j, k, l$ around a plaquette) is a rough measure of $\prod \mathrm{sgn}[(\mathbf{S}_l \cdot \mathbf{S}_i)]$. This equivalence is stronger for larger $J$, since then $\mathbf{S}$ is less likely to take the long way between its values on the lattice sites. Later we will see that the resemblance between the gauge model and that of equations (7,8) is greatest in the limit of small $K$. We have performed Monte Carlo simulations of the two related lattice nematic models of eqns. (7-9) and (11), and our results are discussed in detail in section IX.

## IV. EXACT EQUIVALENCES

In this section, we show that the gauge model must have three distinct phase transitions at the points M, I and H on the phase diagram of figure 3. This provides strong preliminary evidence that there are at least the three phases which actually appear there.

### A. Complete defect suppression: $K = \infty$

When $K = \infty$, only defect-free configurations contribute to the partition function. The product of the $U_{ij}$ around each elementary plaquette, hence around any closed path, then must be $+1$. The defect density is zero everywhere, and the gauge can be chosen [24] so that $U_{ij} = +1$ for each link $(ij)$. Specifically, this can be done as follows: For each site $i$, pick a path of links $P(i)$ from the origin to $i$ and define

$$\sigma_i = \prod_{(kl) \in P(i)} U_{kl}. \tag{12}$$

Since the product of link variables around any closed path is $+1$ at infinite $K$, $\sigma_i$ so defined is independent of the specific choice of path $P(i)$. Furthermore,

$$U_{ij} = \sigma_i \sigma_j, \tag{13}$$

as can be seen by constructing $P(i)$ as the concatenation of $P(j)$ with the single link $(ij)$. Thus, the partition function becomes

$$Z = \sum_{\{\sigma_i\}} \int [d\mathbf{S}] \exp \left\{ -J \sum (\sigma_i \mathbf{S}_i) \cdot (\sigma_j \mathbf{S}_j) \right\}, \tag{14}$$

and the $\sigma$'s embody the freedom of gauge choice. Changing to new variables $\mathbf{S}'_i = \sigma_i \mathbf{S}_i$, each term is independent of the $\sigma$'s and the partition function is $2^N$ times that for an $n = d = 3$ Heisenberg model. Thus for $K = \infty$ there is a second order transition in the three-dimensional Heisenberg universality class at [20] $J \simeq 0.693$.

We show in the next section that this Heisenberg transition persists for large but finite $K$. Finite size scaling analysis of our Monte-Carlo simulations confirms that the transition remains in the Heisenberg universality class out to the multicritical point $M$. Thus "defect fugacity" is irrelevant, in the renormalization group sense.

For complete defect suppression ($K = \infty$), in the gauge with all $U_{ij}$ equal to $+1$, the ordered phase at large $J$ is characterized by a non-zero total magnetization. One may instead check for a non-zero limit of the usual spin-spin correlation function $\lim_{|i-j|\to\infty}\langle \mathbf{S}_i \cdot \mathbf{S}_j\rangle > 0$. This is generalized to arbitrary gauge in the next section.

### B. Pure gauge theory: $J = 0$

For $J = 0$ the spins decouple completely, and our model (eqn. 11) reduces to the three-dimensional pure Ising lattice gauge theory on a cubic lattice. This gauge theory can be mapped onto the ordinary three-dimensional Ising spin model by a duality transformation [25]. The critical coupling $J_c = 0.22165 \pm 0.00003$ for the three dimensional Ising spin model (from Monte Carlo [26] and high-temperature series [27] studies) then implies a critical point in the Ising universality class at $K \simeq 0.765$. While $J = 0$, $K \neq 0$ is of course an unphysical limit of our model, the value of the mapping is in its use as a starting point from which we will argue that the transition persists and remains in the Ising universality class, for small $J > 0$. Finite size scaling studies of our Monte-Carlo simulations indicate that this Ising transition persists all the way to the multicritical point $M$ of figure 3.

### C. No defect suppression: $K = 0$

When $K = 0$, the Ising link variables on different links are decoupled from one another. It is then trivial to trace over $U_{ij} = \pm 1$ and obtain an effective Hamiltonian for the spins alone (subtracting a constant $N \log 2$):

$$H_{\text{eff}}[\mathbf{S}_i] = \sum_{\langle ij\rangle} \ln\left\{\sum_{U_{ij}} \exp(JU_{ij}\mathbf{S}_i \cdot \mathbf{S}_j)\right\} - N\log 2 = \sum_{\langle ij\rangle} \ln\cosh(J\mathbf{S}_i \cdot \mathbf{S}_j) = \sum_{\langle ij\rangle} \frac{1}{2}J^2(\mathbf{S}_i \cdot \mathbf{S}_j)^2 + \mathcal{O}(J^4). \quad (15)$$

Local gauge invariance of eqn. (11) guarantees that tracing over the $U$'s will generate an effective Hamiltonian for the spins which is even in each $\mathbf{S}$ separately, *i.e.* which has local inversion symmetry.

Like the Hamiltonian (eqn. 7) of the Lebwohl-Lasher model, $H_{\text{eff}}$ is a function of $(\mathbf{S}_i \cdot \mathbf{S}_j)^2$; the resemblance is even stronger when it is Taylor expanded in $J$. Thus it is not surprising that, also like the Lebwohl-Lasher model, it has a single, first order phase transition between a nematic and an isotropic phase, indicated by the point N in figure 3 at $J \approx 1.9$.

## V. PHASES OF THE NEMATIC GAUGE THEORY

In this section, we describe the macroscopic distinction among the three phases depicted in figure 3. In the nematically ordered phase $N$, rotational symmetry breaking and nonzero order parameter are accompanied by a non-zero helicity modulus [28], or spin-wave stiffness. In our model, the one-Frank-constant approximation is exact. In a more general model than ours, which included space-spin coupling, the single spin-wave stiffness is replaced by the three Frank constants. This means that if the nematic fluid is confined to a box of side $L$, imposition of boundary conditions with a relative angle $\theta < \pi$ between the directors on opposite faces of the box raises the free energy density by $\Upsilon(\theta/L)^2$ over its value for periodic boundary conditions (*i.e.,* $\theta = 0$). The helicity modulus $\Upsilon$ is positive in the ordered phase and increases with the degree of ordering. Note that for fixed $\theta$, the difference in free energy *density* vanishes in the thermodynamic limit $L \to \infty$. In the one-Frank-constant approximation, $\Upsilon = K/2$, where $K$ is the Frank constant.

An analogous measure of the free energy cost of disclination lines can also be constructed *via* response to changes of boundary conditions. Imagine a cylinder of height $L$ and radius $R$ filled with a nematic fluid. Consider two boundary conditions:

(a) No disclinations are permitted to pierce the boundary, so that all defects must form loops contained entirely within the cylinder.

(b) A single disclination is forced through the centers of the upper and lower faces of the cylinder, but no other defect lines are allowed to pierce the surface. Internal disclination loops coexist with the externally imposed defect.

The three phases appearing in figure 3 are distinguished by the dependence of the free energy difference $\delta\mathcal{F}$ between boundary conditions (a) and (b) on the radius and height of the cylinder. Note that, as with the manipulation of boundary conditions used to measure the helicity modulus, this free energy difference is not extensive.

**Nematic phase.** In the *nematically ordered phase N*, the extra defect imposes a variation of the director arbitrarily far from the core, since the director must undergo a net rotation of $\pi$ along any path encircling the defect.

The free energy difference between the two boundary conditions is therefore governed by the helicity modulus, *viz.*,

$$\begin{aligned}\delta F_N &\geq \Upsilon \int \left(\frac{\pi}{2\pi r}\right) 2\pi r \,\mathrm{d}r \\ &= C_N \; L \ln(R/a),\end{aligned} \tag{16}$$

where $C_N(J,K) = \pi\Upsilon/2$ is the long wavelength nematic stiffness and $a$ is the defect core size. The defect line tension $\delta F/L$ is therefore logarithmically divergent with system size whenever long-range nematic order exists, and vice versa.

A similar calculation shows that the interaction energy between a pair of externally imposed defects has a logarithmic dependence on separation. Thermal fluctuations generate spontaneous defect loops within the cylinder, but these loops have a strong energetic preference to avoid director variation far from their cores. They can achieve this by being small or binding in pairs. In this phase, the defects can be said to be *confined.*

For small enough $J$ (that is, $J$ below a $K$-dependent threshold $J_c(K)$, the long-wavelength nematic stiffness vanishes and the free energy difference between (a) and (b) is no longer logarithmic with $R$. Depending on the degree of defect suppression, however, there are two distinct regimes which correspond to the two isotropic phases $I$ and $T$ of our model.

**Topologically ordered phase.** Let us consider a state without long-range nematic order but with sufficient short-range order to permit the coarse-graining leading up to the gauge model of eqn. (11). If the defect core energy $K$ is sufficiently large (*e.g.,* $K=\infty$), then the local molecular axes can be consistently assigned a continuous orientation throughout the system *even in the absence of long-range order.* This assignment effectively converts the short-range ordered nematic director field into a short-range-ordered vector field. We call this novel phase the topologically ordered state. Its physical properties are isotropic (since it lacks long range order), but it is nevertheless distinct from the usual isotropic phase. This distinction is quantified by different asymptotic behaviors of the defect free energy.

Despite the absence of long-range nematic order in the large-$K$ disordered phase $T$, the free energy cost per unit length of disclination remains nonzero. The bending imposed on the spins by the presence of the defect is screened out over the correlation length, since the helicity modulus is zero. Near the defect core, however, the extremely tight bending (along with the bare defect core energy $K$) produces a non-vanishing defect line tension which is independent of the radius of the cylinder. The free energy difference between the two boundary conditions is given by

$$\delta F_T \sim C_T \; L, \tag{17}$$

where $C_T(J,K)$ is a constant which depends on the bare core energy and the elastic energy within a nematic correlation length of the core.

In contrast with the nematic phase, in the topologically ordered state there is no long-range interaction between disclinations mediated by the nematic fluid, since long range nematic order is not present. Defects are therefore *unbound*, and loops of finite extent will proliferate. Viewed from far away compared with their linear extent, these loops are either topologically trivial or the point defects of the nematic (hedgehogs). The picture of this phase as a gas of unbound hedgehogs is similar to that for the ($d=n=3$) Heisenberg model in reference [18].

If the nematic interaction $J$ is now increased, nematic order will develop from a state without topological defects. Since the relevant configurations are those of a vector field, an ordering transition in the Heisenberg universality class is expected. This is discussed more carefully in sections IV A and VII C. If the renormalized defect fugacity is small enough that arbitrarily large defect loops do not proliferate, the transition will be in the Heisenberg class along the entire line MH, as confirmed below by Monte Carlo studies.

**Isotropic phase.** As the core energy $K$ is diminished at fixed, weak nematic interaction $J$, thermal fluctuations will cause the defect line imposed by boundary condition (b) to meander through the system. The system will then gain an entropy proportional to its length $L$, and the defect line tension will diminish. At a critical point the line tension will vanish, and the free energy difference between (b) and (a) will become independent of the dimensions of the cylinder for large cylinders:

$$\mathcal{F}_I = C_I(J,K). \tag{18}$$

We will show below that this transition can be understood as an Ising lattice gauge theory whose critical point lies in the universality class of the three-dimensional Ising model.

In the resulting disordered phase $I$ the nematic stiffness and the defect line tension both vanish, so the free energy difference between the two boundary conditions remains finite as $L$ and $R$ tend to infinity. In this phase there is a nonzero density of infinitely long defects (see figure 5b), which can be called a *condensation* of disclinations.

## VI. ORDER PARAMETERS

An order parameter for a system with a local gauge symmetry must be gauge-invariant [29]. In our model the gauge group is $\mathbb{Z}_2$ and the local gauge transformation is given by eqn. (10). A product of spins and link variables is therefore gauge-invariant if the number of factors ($\mathbf{S}_i$ or $U_{ij}$) pertaining to each site $i$ is even. Observables containing only gauge fields are therefore made of Wilson loops — products of the link variables around a closed curve $\gamma$ (typically a rectangle):

$$W(\gamma) \equiv \prod_{\langle kl\rangle \in \gamma} U_{kl}. \tag{19}$$

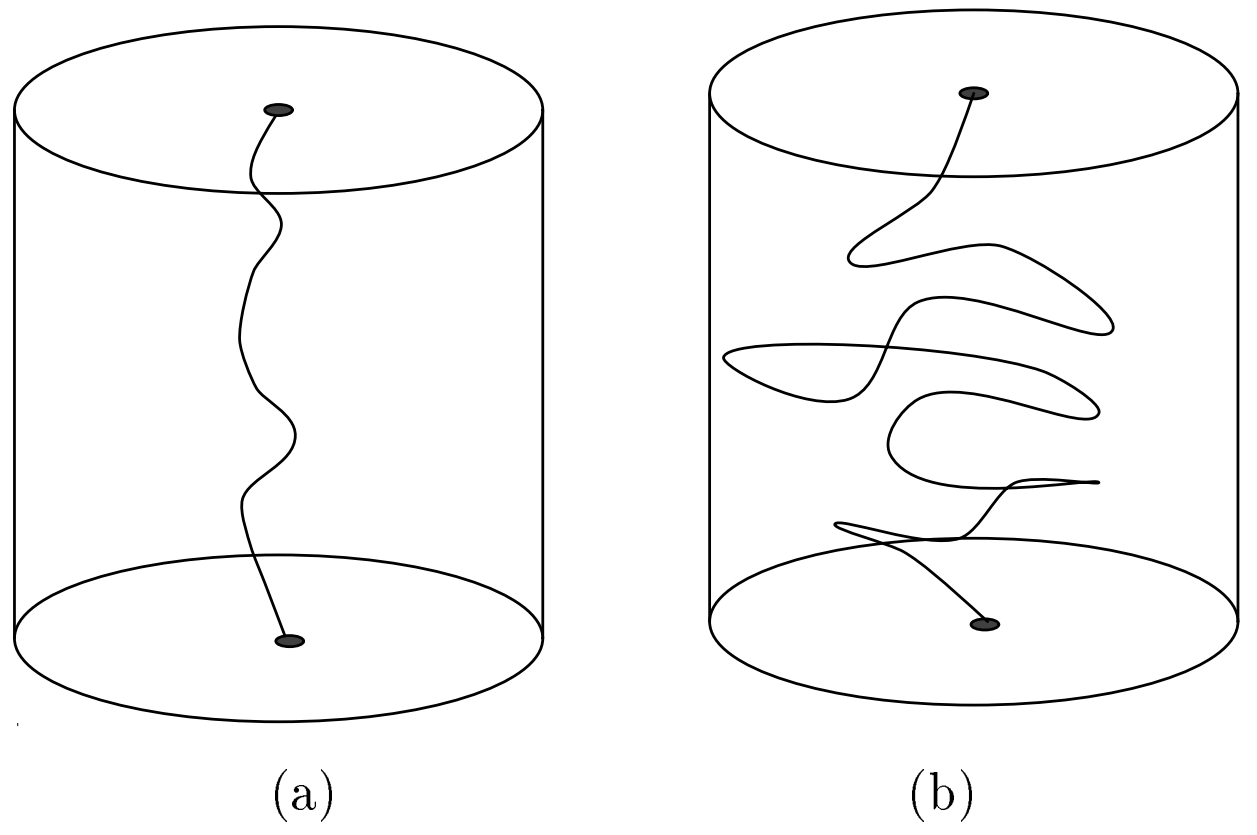

FIG. 5. A caricature of the defect which is forced by boundary conditions to traverse the system in (a) the $N$ or $T$ phase and (b) the $I$ phase.

Purely spin-dependent quantities are gauge invariant if each spin enters an even number of times, such as the the defect counting term in eqn. (8), and the familiar traceless, symmetric, tensor order parameter

$$Q_{\alpha\beta}(i) = S_{i\alpha}S_{i\beta} - \delta_{\alpha\beta}/3, \tag{20}$$

used in the Landau theory of the nematic-isotropic transition. In the absence of an applied ordering field, long-range nematic order can be detected by calculating the correlation function

$$N(i,j) = \langle \mathrm{Tr}\ \mathrm{Q(i)Q(j)} \rangle = \langle (\mathbf{S}_\mathrm{i} \cdot \mathbf{S}_\mathrm{j})^2 - 1/3 \rangle. \tag{21}$$

$N(i,j)$ is invariant under global spin rotation, and thus measures the tendency of the spins to align along a common axis, regardless of its orientation. When $Q_0^2 \equiv \lim_{|\mathbf{r}_i - \mathbf{r}_j| \to \infty} N(i,j) > 0$, or equivalently $\langle Q(i) \rangle \neq 0$, the system is nematically ordered. Thus, we have a local order parameter to distinguish the nematically ordered phase $N$ from the isotropic phases $I$ and $T$.

Gauge invariant quantities can also be constructed by using both spin and link variables, as in the following path-dependent spin-spin correlation function:

$$C(i,j;\gamma) = \left\langle \mathbf{S}_i \cdot \mathbf{S}_j \prod_{\langle kl \rangle \in \gamma} U_{kl} \right\rangle, \tag{22}$$

with the product of link variables taken along some path $\gamma$ joining sites $i$ and $j$. This is a generalization to arbitrary gauge of the ordinary spin-spin correlation. In the infinite $K$ limit, the choice of path is immaterial, since the product of links around any closed path is guaranteed to be one. Thus, in this limit, $\langle C(i,j;\gamma) \rangle$ is equal to $\langle \mathbf{S}_i \cdot \mathbf{S}_j \rangle$ for the ordinary Heisenberg model at the same value of $J$. For any finite $K$, however, the path-dependent correlator (eqn. 22) always decays to zero exponentially with the separation between $i$ and $j$, independent of the path $\gamma$. This is not really surprising. For finite $K$, the links represent fluctuating degrees of freedom (the topological defects), and a single "weak link" will change the sign of the path-dependent product. (The same phenomenon occurs with the Wilson loop expectation value. Even at large $K$ and in the presence of other ordering, it decays exponentially as a result of a non-zero density of very small defect loops.)

An individual spin is not gauge invariant, so $\langle \mathbf{S} \rangle \neq 0$ is possible only if the gauge is fixed. Nevertheless, global rotation symmetry can be broken by a preferential alignment of the spins along some axis. This is precisely what is measured by $N(i,j)$ (eqn. 21) or $\langle Q(i) \rangle$ (eqn. 20). Ferromagnetic ordering implies nematic ordering but the converse is false. That the global $SO(3)$ symmetry actually is spontaneously broken at some finite, non-zero value of $J$ for any $K$ is clear from the results of the previous section. We know that it is broken at finite $J$ (points N and H on figure 3) in both of the limits $K = \infty$ and $K = 0$. Reducing $K$ introduces frustration and makes ordering more difficult. The only reasonable conclusion is that there is a line $J_c(K)$, monotonically decreasing with $K$, at which nematic ordering occurs.

For the pure Ising gauge theory ($J = 0$), the two phases on either side of the Ising transition discussed in section IV B above are distinguished by the asymptotic behavior of large Wilson loops $\langle W(L,T) \rangle$, for which the path $\gamma$ is a closed rectangular loop of sides $L$ and $T$. In the small $K$ phase, the Wilson loop follows an "area law" for sufficiently large $L$ and $T$: $\ln\langle W(L,T) \rangle$ is proportional to the area $LT$ of the closed path. This corresponds to the "confining" phase of the gauge theory. In the large-$K$ phase, on the other hand, sufficiently large Wilson loops obey a "perimeter law," so that $\ln\langle W(L,T) \rangle$ is proportional to the perimeter $L + T$ of the closed path. This corresponds to the "free-charge" phase of the gauge theory. These results emerge from expansion methods described further in section VII A. As also shown there, the "defect line tension" $\delta\mathcal{F}/L$ is related by duality to the spin-spin correlation function of the Ising spin model.

The distinction between the fully disordered and topologically ordered isotropic phases is more subtle when $J$ is nonzero. There is no local order parameter which distinguishes between the $I$ and $T$ phases. The line tension is therefore quite important because it does perform this function. Coupling of spin and gauge degrees of freedom results in a decay of Wilson loops which is asymptotically a perimeter law for any nonzero $J$, with a crossover scale of $L_\times = 4\log(J/3)/\log(\tanh K)$. The defect line tension described in section V, however, remains a valid diagnostic of topological order.

The results discussed so far show quite clearly that there is a line connecting the first-order transition at N to the second-order one at H as depicted in figure 3, at which global spin rotational symmetry is broken. there is also a critical point at I. We have until now somewhat implicitly assumed that this point actually separates two completely distinct phases $I$ and $T$, though the line tension appears to be a good diagnostic of this distinction, and offers strong support for it. Accepting this point (which is established more carefully in later sections), the simplest possible phase diagram topology, *i.e.* that with the fewest number of phases, is shown in figure 3, which in fact is the outcome of our Monte Carlo simulations.

In section VII, we demonstrate analytically that the continuous Ising transition near I persists for nonzero $J$, and the Heisenberg transition near H persists for finite defect activity $e^{-2K}$. In section IX we fill in the analytically intractable interior of the phase diagram using Monte Carlo simulation. This work shows that the three transition lines emanating from the border of the phase diagram meet at a multicritical point M. The jump of the order parameter at the symmetry-breaking transition goes to zero at M, so that the transition is first order to the left (smaller $K$) of M and continuous to the right. Finite size scaling has been employed to verify that the entire continuous transition line is in the Heisenberg universality class, not just the point at infinite $K$. Calculations of the specific heat strongly suggest the existence of the Ising transition line originating from the pure gauge theory transition. This is verified by calculations of the defect line tension.

## VII. ROBUSTNESS OF CONTINUOUS TRANSITIONS

In this section, we demonstrate that the continuous Heisenberg $N/T$ and Ising $T/I$ transitions found in section IV at $K=\infty$ and $J=0$ respectively, persist for $K<\infty$ and $J>0$. These transitions are therefore generic and occur for a finite range of material parameters, and not just at isolated points. Our tool is perturbation theory in $e^{-K}$ and $J$, which can be developed in terms of polymer expansions. The general formalism of such expansions is developed in subsection VII A. Subsections VII B and VII C treat the limit $K\to\infty$, while subsections VII D and VII E deal with the limit $J\to 0$. Finally, we treat the limit $K\to 0$ in subsection VII F.

### A. Polymer Expansions

Polymer expansions for lattice models express the partition function and correlation functions in terms of an interacting system of self-avoiding chains of links on the lattice. The simplest example is the familiar high-temperature expansion of the Ising model; Mayer expansions may also be viewed as polymer expansions. For more details on this subject, the derivation of eqn. (30), and careful discussions of the convergence question, the reader is directed to references [30–32].

With Ising variables ($\sigma=\pm 1$), the simple identity

$$e^{x\sigma}=\cosh(x)(1+\sigma\tanh(x)), \tag{23}$$

is remarkably useful. By taking $\sigma=U_{ij}$ and $x=J\mathbf{S}_i\cdot\mathbf{S}_j$, this identity can be applied to the part of the Boltzmann factor arising from the spin term of our Hamiltonian, eqn. (11). Expansion of the ensuing product produces a sum of terms, each of which contains factors $U_{ij}\tanh(J\mathbf{S}_i\cdot\mathbf{S}_j)$ for the links $(ij)$ in a distinct set. We decompose each such set into constituent "polymers" connected at the lattice sites, and define the activity of a polymer $\omega$ as

$$\rho(\omega)=\int\prod_{i\in\omega}d\mathbf{S}_i\,\rho(\omega,\mathbf{S}) \tag{24}$$

where $d\mathbf{S}_i$ denotes the usual integration measure over the directions of $\mathbf{S}_i$ (normalized: $\int d\mathbf{S}=1$), and

$$\rho(\omega,\mathbf{S})=\prod_{\langle jk\rangle\in\omega}\tanh(J\mathbf{S}_j\mathbf{S}_k) \tag{25}$$

is a quantity we will need later when discussing small $K$. A polymer clearly has zero activity unless it is closed, *i.e.* an even number of constituent links impinge on each lattice site. (When calculating a correlation function instead of the partition function, we wil alter this definition slightly, so that some open polymers may have nonzero activity.) In this representation, the complete partition function is written

$$Z(J,K)=\sum_C Z(K)\langle W(C)\rangle_G\prod_{\omega\in C}\rho(\omega), \tag{26}$$

where the sum runs over collections $C$ of non-intersecting spin polymers, and $W(C)$ is the associated generalized Wilson loop, a product of factors $U_{ij}$ for each link in $C$. Its expectation, evaluated here in the *pure gauge model* at coupling $K$ (as indicated by $\langle\rangle_G$), can also be written as

$$\langle W(C)\rangle_G=\frac{Z_C(K)}{Z(K)}, \tag{27}$$

where $Z(K)$ is the gauge theory partition function at coupling $K$, and $Z_C(K)$ the partition function in the presence of "source" $C$.

Spin polymers interact via both the gauge field and a hard core repulsion.

### B. Large $K$

Now we evaluate $\langle W(C)\rangle_G$ for some fixed $C$, by introducing a second type of polymer. At large $K$, the gauge field configurations are most conveniently represented in terms of defect loops on the dual lattice. A link on the dual lattice pierces a unique plaquette on the original lattice. For a configuration of $\{U_{ij}\}$, that (dual) link is part of the defect network if the product of $U_{ij}$'s around its associated plaquette is $-1$. As with the spin polymers, we decompose the defect network into pieces which are connected at the dual-lattice sites. We call the pieces "defect loops," although the nomenclature is not ideal since such a loop may cross itself many times.

A defect loop $\gamma$ of total length $|\gamma|$ has a $C$–dependent activity given by

$$z(\gamma, C) = (-1)^{i(\gamma,C)} e^{-2K|\gamma|}, \tag{28}$$

where $i(\gamma, C) = \pm 1$ is the parity of the linking of $\gamma$ with $C$, which is equal to the product of the linking parities of $\gamma$ with the separate constituents of $C$. (For two loops $\gamma$ and $\omega$, $i(\gamma, \omega)$ is $-1$ if $\gamma$ wraps around $\omega$ an odd number of times, and $+1$ if an even number).

By construction, the defect loops do not overlap. We may also think of this as arising from a hard-core repulsion. The partition function in the presence of source $C$ will then be written as

$$Z_C(K) = \sum_D \prod_{\gamma \in D} z(\gamma, C) \prod_{\gamma,\gamma' \in D} (1 - l(\gamma, \gamma')), \tag{29}$$

where $l(\gamma, \gamma')$ is $+1$ if the defects $\gamma$ and $\gamma'$ overlap, and is 0 if they don't. The sum over collections $D$ of defects does not then need to be restricted to non-overlapping sets. Each term in the expansion of the product of factors $(1 - l(\gamma, \gamma'))$ can be associated to a graph whose nodes represent the polymers $\gamma$, and in which two nodes corresponding to $\gamma$ and $\gamma'$ are joined by a line if they overlap, *i.e.* if $l(\gamma, \gamma') = 1$.

This formulation is advantageous when we pass to the logarithm, *i.e.* the free energy. Explicitly,

$$F_C(K) = -\ln Z_C(K) = {\sum_D}' \frac{n(D)}{|D|\,!} \prod_{\gamma \in D} z(\gamma), \tag{30}$$

where the sum is over connected graphs $D$ containing $|D|$ nodes. The "index" of $D$ is defined as

$$n(D) \equiv {\sum_{G \subset D}}' (-1)^l (G), \tag{31}$$

where the sum is over connected subgraphs $G \subset D$ which contain *all* the nodes of $D$ (thus $|G| = |D|$), and $l(G)$ is the number of lines in $G$. The formula (30) for the free energy is an expansion (cf. eqn. 28) in powers of

$$\alpha \equiv e^{-2K}. \tag{32}$$

For $\alpha$ sufficiently small ($K$ sufficiently large), the combined influence of factors of $\alpha$ and the requirement of connectedness suppresses the higher-order contributions enough that we can prove convergence of the expansion.

We can now see that $\ln\langle W(C)\rangle_G = \ln Z_C(K) - \ln Z(K)$ is a difference of two expansions like eqn. (30), one with activities $z(\gamma, C)$ and the other with $z(\gamma)$. Only terms containing polymers with $z(\gamma, C) \neq z(\gamma)$ need to be calculated; all others are killed by the subtraction. To see this in action, we calculate the contribution of the smallest defect loops to $\langle W(C)\rangle_G$. In that case, the $D$'s occurring in the sum in eqn. (30) consist of single defect loops of four links going around the perimeters of elementary plaquettes on the dual lattice. Only if $\gamma$ encircles $C$ are $z(\gamma, C) = -z(\gamma)$ $z(\gamma) = \alpha^4$ different. The result is therefore

$$\ln\langle W(C)\rangle_G = -2e^{-8K} \cdot |C| + \mathcal{O}(e^{-12K}). \tag{33}$$

This result can now be used to derive an effective Hamiltonian for the spins alone. We sum over the defect configurations to express $Z(J, K)$ purely in terms of spin-polymers. In the expansion of eqn. (26), each factor $\tanh(J\mathbf{S}_i \cdot \mathbf{S}_j)$ in a spin-polymer activity $\rho$ (eqn. 24) is accompanied by a factor of $U_{ij}$, which goes into $W(C)$. From eqn. (33), we can think of each link in $C$ as bringing a factor $\exp(-2\alpha^4)$ to the expectation $\langle W(C)\rangle$. Alternatively, we can put this factor with the tanh to write, correct to order $\alpha^4$,

$$Z(J, K) = \int \prod_i d\mathbf{S}_i \prod_{\langle jk \rangle} \cosh(J\mathbf{S}_j \cdot \mathbf{S}_k) \times \left(1 + e^{-2\alpha^4} \tanh(J\mathbf{S}_j \cdot \mathbf{S}_k)\right). \tag{34}$$

As remarked before eqn. (24), the constraint to closed loops is still in force by virtue of the integration over the spins.

Taking a logarithm gives an effective Hamiltonian

$$H_{\text{eff}} = \sum_{\langle ij \rangle} \Big\{ (1 - 2\alpha^4) J\mathbf{S}_i \cdot \mathbf{S}_j + \alpha^4 \sum_{n=2}^{\infty} \frac{(-2J\mathbf{S}_i \cdot \mathbf{S}_j)^n}{n!} \Big\} + \mathcal{O}(\alpha^6). \tag{35}$$

To this order, the effective Heisenberg coupling is reduced to $J_{\text{eff}} = (1 - 2\alpha^4)J$. Since eqn. (35) contains additional terms higher order in the spins but the same order in $\alpha$, we can say $J_c(\alpha) - J_c(\alpha = 0) = \mathcal{O}(\alpha^4)$, but cannot predict the numerical prefactor.

The effective Hamiltonian in eqn. (35) gives the same free energy as the original model. If one wants to compute expectations, it is helpful first to restore gauge invariance by multiplying $\mathbf{S}_i \cdot \mathbf{S}_j$ by $U_{ij}$, where these link variables are subject to an infinite effective $K$. Going through the same arguments, one finds that expectations of quantities involving link variables $U_{ij}$ (the path dependent scalar product $C(i, j; \gamma)$ of eqn. 22, for instance ) can be computed by replacing $U_{ij}$ inside the expectation with $e^{-2\alpha^4} U_{ij}$ and then using $H_{\text{eff}}$. Thus, exponential decay of $C(i, j; \gamma)$ is maintained even in the ordered phase.

The program of eliminating the link variables $U$ appears to be going well, and the result (eqn. 35) argues for suggestive of the irrelevance of small defect activity. However, this is the lowest order result; only the smallest defects have been eliminated. Higher order (in $\alpha$) terms, while individually small and formally irrelevant, proliferate alarmingly at higher order. It is not entirely obvious that they can be neglected, or that the "sum" even exists. In section VII C below, the issue of defect activity irrelevance is taken up again, within a renormalization group approach.

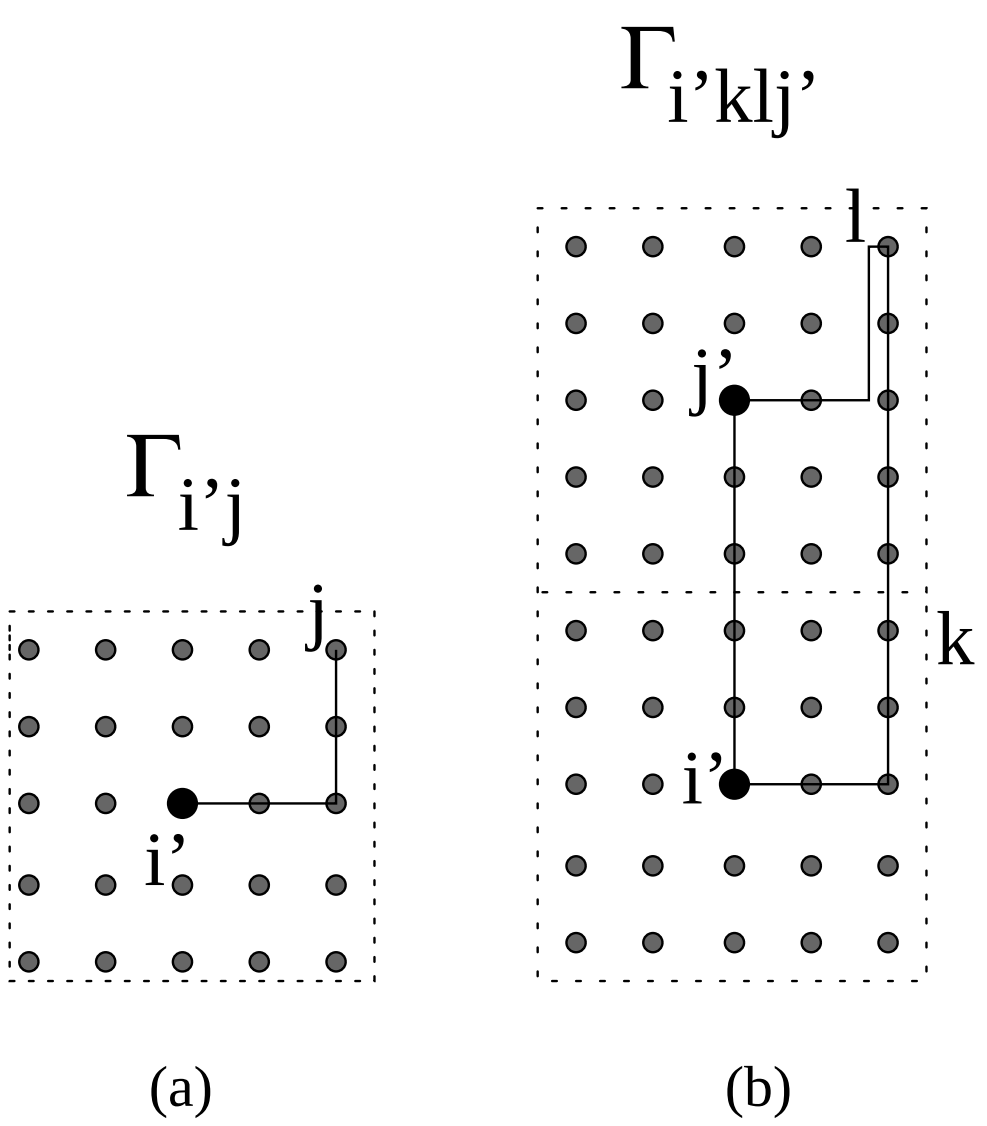

FIG. 6. The lattice paths which occur in the Real-Space blocking scheme for (a) the spins and (b) gauge (link) fields, in the case of two dimensions.

### C. Irrelevance of Defect Activity

The first task in developing a real-space renormalization group approach is to define block variables. We need to block both spins and link variables. The spins can be handled in a usual way, with a bit of care to insure gauge invariance. The link variables are not so familiar in this context and require some thought. We will present one particular blocking scheme, and then comment on the motivation.

Figure 6 illustrates the blocking procedure. Sites in the blocked lattice are denoted with a prime superscript, thus, site $i'$ is associated with the cubical block $B(i')$ of sites centered on $i'$ with side length $2L+1$:

$$B(i') \equiv \{j \,:\, |x_\alpha(j) - x_\alpha(i')| \le L,\, \alpha = 1,2,3\}. \quad (36)$$

The contribution of the spin at $j$ to the block spin $\mathbf{S}'_{i'}$ is found by parallel transporting it to $i'$, using some set of standard paths. A convenient choice, illustrated in fig. 6 for the case of two dimensions, is to move along coordinate directions in reverse lexicographic order, *i.e.* move along the z-direction until the z-coordinate equals that of the destination, then along the y-direction, then along the x-direction. Denote this path from $j$ to $i'$ by $\Gamma_{i',j}$. The blocked spin is now defined as

$$\mathbf{S}'_{i'} \equiv \frac{\zeta_S}{L^3} \sum U(\Gamma_{i',j}) \mathbf{S}_j, \quad (37)$$

where the parallel transport factors are

$$U(\Gamma_{i',j}) \equiv \prod_{kl \in \Gamma_{i',j}} U_{kl}. \quad (38)$$

The rescaling factor $\zeta_S$ will be adjusted to keep the distribution centered around vectors of unit length.

For two neighboring sites $i'$ and $j'$ in the blocked lattice, we might try to make a blocked link variable by simply taking the product of link variables along the straight path from $i'$ to $j'$ denoted by $\tilde{\Gamma}_{i',j'}$. This is a really a decimation scheme. It turns out to be much better to take products of link variables along paths which are deformations of this straight path on a scale $L$ and average them. Defining closed loops

$$\Gamma_{i',k,l,j'} = \Gamma_{i',k} \cup \tilde{\Gamma}_{k,l} \cup \Gamma_{j',l} \cup \tilde{\Gamma}_{i',i'}, \quad (39)$$

our blocked link variable is (see figure 6)

$$U'_{i'j'} \equiv \zeta_U\, U(\tilde{\Gamma}_{i',j'}) \frac{1}{L^3} {\sum_{k,l}}' U(\Gamma_{i',k,l,j'}). \quad (40)$$

The prime on the sum denotes restriction to $k \in B(i')$, $l \in B(j')$. Again, there is a rescaling factor to keep the link-magnitude distribution centered at one.

The preservation of gauge invariance by our blocking scheme is easily checked. Performing the gauge transformation $\mathbf{S}_i \to \phi_i \mathbf{S}_i$, $U_{ij} \to \phi_i \phi_j U_{ij}$ on the original lattice first ($\phi_i = \pm 1$ chosen independently for each $i$), and then blocking results in

$$\begin{aligned} \mathbf{S}'_{i'} &\to \phi_{i'} \mathbf{S}'_{i'} \\ U'_{i'j'} &\to \phi_{i'} \phi_{j'} U'_{i'j'}. \end{aligned} \quad (41)$$

This is a gauge transformation on the block lattice, as required.

After $m$ renormalization group steps, the running Hamiltonian looks like

$$\begin{aligned} H_m = &\sum \Phi_m(\mathbf{S}_i^2) + \sum \Phi_m^U(U_{ij}^2) \\ &+ J_m \sum U_{ij} \mathbf{S}_i \cdot \mathbf{S}_j + K_m \prod U_{ij} U_{jk} U_{kl} U_{li} + \cdots, \end{aligned} \quad (42)$$

and we need to determine the flow of the various couplings. The first two terms provide the weighting for the magnitudes of the spins and link variables and their indicated functional dependence is dictated by gauge invariance and rotational symmetry. On the original lattice ($m = 0$), they were delta functions at one. The ellipsis indicates all other interactions generated by the renormalization group transformation.

Our blocking scheme treats large and small defect loops differently, and this is reflected in the structure of $H_m$. Defects smaller than the current lattice spacing $L^m$ are effectively integrated out in the sense that they no longer appear as extended objects, but they suppress the magnitudes of the blocked links. Their effects are reflected in $\Phi^U$. After blocking, but before rescaling the link variables, $U^2_{i'j'}$ typically has a value $\sim e^{-8L\alpha}$, from the small defects. ($\alpha = e^{-2K}$) This has an influence of the magnitude of the product $\prod U_{i'j'} U_{j'k'} U_{k'l'} U_{l'i'}$ around a blocked plaquette. That product, like the blocked link variables $U_{i'j'}$ making it up, is an average over many paths. It will not be negative unless most of them are encircled by a defect. This can be accomplished by one large defect of length $\sim L$, or an even greater length of small ones.

The value of $K_m$ is therefore determined by the probability to have such a large length of defect ($\sim L^m$ in original lattice units) through the blocked plaquette; everything else is absorbed into $\Phi^U$. As in the calculation of $K_{\text{eff}}$ in section VII D, the spins also contribute to the renormalization of $K$, and tend to suppress it. Let us neglect this, though, and make the conservative estimate

$$K_{m+1} \sim L(K_m - c)e^{-16L\alpha_m}, \tag{43}$$

where the $c$ comes from an entropy factor (the number of distinct defects of total length $L$ is $\sim \exp cL$). The rescaling of the link variables has also been taken into account in making this estimate and accounts for the final exponential factor.

As long as the initial value of $K$ is large enough, and we don't try to take too large blocking steps, the defect fugacity $\alpha$ is being rapidly driven to zero, exponentially fast, in fact. If $K$ starts out too small, the entropy dominates, and this scenario no longer holds [33].

Thus, in order to conclude that a small defect activity does not alter the Heisenberg universality of the transition, it is only necessary to verify that in the process of vanishing it does not induce any extra interactions which are themselves relevant. The fundamental constraints on the form of these are gauge and rotational invariance. Among terms involving only the gauge variables, there are even more irrelevant multi-plaquette terms, and also the bond weight $\Phi^U$, which ultimately amounts to annealed randomness in the Heisenberg model bond strengths. Two of the least irrelevant pure-spin terms which can occur are $(\mathbf{S}_i \cdot \mathbf{S}_j)^2$ and the spins-around-a-plaquette term $\prod(\mathbf{S}_i \cdot \mathbf{S}_j)$, which will be seen again in section VII F. None of these perturbations is relevant. Thus, the presence of weak defect activity does not alter the (Heisenberg) universality class of the transition.

### D. Small $J$

At small $J$, the spins are a small perturbation to the pure gauge theory. We can try to integrate them out, as we did the gauge field excitations in section VII B, to obtain an effective Hamiltonian for the gauge field, with corrections ordered as a power series in $J$. The polymer expansion is (see 24,26)

$$Z(J,K) = \sum_{\{U_{ij}\}} e^{-\beta H_G} \sum_C \prod_{\omega \in C} \left( \rho(\omega) \prod_{\langle ij \rangle \in \omega} U_{ij} \right). \tag{44}$$

At leading order, the only spin polymers occurring are those which go around an elementary plaquette on the lattice, just as with the defect loops on the dual lattice in the $K \to \infty$ limit. The required activity is

$$\begin{aligned} \rho &= \int [d\mathbf{S}] \prod_{\langle ij \rangle \in P} \tanh(J\mathbf{S}_i \cdot \mathbf{S}_j) \\ &= 3(J/3)^4 + \mathcal{O}(J^6), \end{aligned} \tag{45}$$

where the product is over the links around the plaquette $P$. Exponentiating this result gives an effective action for the gauge field [34] with $K_{\text{eff}} = K + J^4/27 + \mathcal{O}(J^6)$.

Terms of higher order in $J$ also introduces gauge couplings on larger closed loops of links which decrease exponentially with the length of loop.

The polymer expansion for the spin problem is known to be convergent for small enough $J$, so the only question is whether it can legitimately be rewritten as an effective action for the gauge field. The entire calculation is very similar to the one performed for the large $K$ regime in section VII B, with the defect loops replaced by spin-polymers. In fact, if our spins were Ising spins, there would be a perfect duality. We believe that the near-duality makes this effective action calculation as solid as the previous one, and establish the robustness of the Ising transition for $J > 0$.

### E. Line Tension Again

The defect-line tension was discussed heuristically in section V. Additional insight can be gained by using the duality between the three-dimensional Ising spin and gauge models. This duality amounts to no more than the banal observation that the polymer representations of the ordinary nearest-neighbor Ising model $H = -J\sum \sigma_i\sigma_j$ and the gauge model in eqn. 11 are *identical* (up to an innocuous overall factor of $(\cosh J)^{3N}$) when the couplings are related by

$$e^{-2K} = \tanh J. \tag{46}$$

The self-duality of the $\mathbb{Z}_2$ gauge-Higgs model (*i.e.* spins and links are Ising variables) is just as clear. Referring to section VII B, we see that the spin polymers and defect loops have identical activities apart from a linking-number related interaction, which is completely symmetric between the two types.

Dual representations of correlation functions are also easily written down by going through the polymer representation. For instance, $\langle \sigma_i\sigma_j \rangle \cdot Z$ corresponds to polymer configurations having one polymer with an odd number of links connected to sites $i$ and $j$. One way to generate this set is to take one fixed configuration $\Sigma$ which satisfies the condition, a single path of links between the two sites for instance, and then forming the symmetric difference with each configuration for $Z$, which are made of closed loops. (The symmetric difference of two sets $A$ and $B$ is $A\Delta B = (A \cup B) \setminus (A \cap B)$.)

In the gauge model, the set of links dual to plaquettes on which the field strength is $-1$ is required to be closed by the nature of the underlying fields. If the couplings are not all the same, the set of plaquettes on which $K_P U_P$, *i.e.* on which the configuration is an excitation over the local ground state is not required to be closed. The defect loops must really be identified with the excited plaquettes, so if we reverse the sign of the coupling on plaquettes in $\Sigma^*$, with $\Sigma$ as above, the partition function of the resulting model is the dual representation of $Z \langle \sigma_i \sigma_j \rangle$. We get exactly the same thing by evaluating the expectation value of the operator

$$D(\Sigma^*) = \prod_{P \in \Sigma^*} \exp(-2K\, U_P) \tag{47}$$

with the original Hamiltonian. This disorder operator, (a 't Hooft operator [35] for our particular gauge group) is therefore the dual representation of the operator $\sigma_i \sigma_j$. The number of defect lines entering or leaving any elementary cube of the lattice has the same parity as the number of plaquettes on that cube for which the coupling has been reversed in sign.

The 't Hooft disorder operator can detect the transition in the pure gauge theory ($J = 0$). As $i$ and $j$ become infinitely separated, $\langle D(\Sigma^*) \rangle$ tends to a non-zero constant at small $K$ and decays exponentially in $|i - j|$ for large $K$. At $J = 0$, the only relevant aspect of the path $\Sigma$ is its endpoints. For $J > 0$, this is no longer true, since the spins are sensitive to the signs of $U_P$ and not just the products $K_P U_P$. As a result, for $J > 0$, $\langle D(\Sigma^*) \rangle$ decays exponentially in the length of the path $|\Sigma|$ for any value of $K$. This behavior is reminiscent of the path-dependent spin-spin correlation function (eqn. 22). This is not an accident; in the case of $\mathbb{Z}_2$ spins, the two are dual to each other.

The defect line tension is similar to the 't Hooft operator, and equivalent at $J = 0$, where the the tension is therefore positive for $K > K_c$ and zero for $K < K_c$. Unlike the Wilson loop or 't Hooft operator, however, the line tension continues to be a good diagnostic for nonzero $J$. Rigorous results [36] show that it vanishes in some region around $J = K = 0$ and is strictly positive in some region around $K = \infty$, $J = 0$ in our phase diagram. Whether these behaviors hold throughout the entire $I$ and $T$ phases however, is not rigorously established.

To go further, consider the truncated energy-energy correlation

$$\langle \epsilon(ij); \epsilon(kl) \rangle = \langle \sigma_i \sigma_j \sigma_l \sigma_m \rangle - \langle \sigma_i \sigma_j \rangle \langle \sigma_l \sigma_m \rangle, \tag{48}$$

for nearest neighbor pairs $(i, j)$ and $(l, m)$, which yields the specific heat when summed over links $(lm)$. Choose $\Sigma$ (as in the spin-spin correlation construction above) to be the two plaquettes dual to the links $(ij)$ and $(lm)$, so that the dual representation of $\langle \epsilon(ij)\, \epsilon(kl) \rangle$ is obtained by weighting each configuration of defects contributing to the gauge theory partition function by an extra factor of $e^{-2K}$ for each of those links which does not occur in the defect set and $e^{+2K}$ for each which does, and similarly for $\langle \epsilon(ij) \rangle \langle \epsilon(kl) \rangle$.

We use the notation $\chi(ij) \equiv U_{(ij)^*}$, for the product of link variables around the plaquette $(ij)^*$ which is dual to the link $(ij)$. Thus, $\chi(ij) = 1$ if dual link $(ij)$ is in the defect set and $\chi(ij) = 0$ otherwise. Then we have the correspondence

$$\begin{aligned} &\langle \epsilon(ij); \epsilon(kl) \rangle \\ &= 4 \sinh^2 2K \langle \chi(ij); \chi(kl) \rangle \\ &= \sinh^2 2K \langle U_{(ij)^*}; U_{(kl)^*} \rangle \end{aligned} \tag{49}$$

The divergence of the specific heat in the spin model shows that this function becomes long-ranged at the critical point. In the gauge theory language, what eqn. (49) measures is the degree to which frustrations of widely separated plaquettes are not independent. We argue that its asymptotic behavior reflects that of the probability for two widely separated points to be in the *same* defect cluster. Multiplying eqn. (49) through by a factor of the partition function $Z$, we rewrite the result as a sum over the defect cluster [37], $\gamma$, which contains dual link $(ij)$, since only such configurations contribute. We will also extract an explicit factor $z(\gamma) = \alpha^{|A|}$ of the weight of cluster $\gamma$ (eqn. 26 with $C = \emptyset$). Thus we write

$$\sum_{\gamma}{}' z(\gamma) \left(1 - \langle \chi(kl) \rangle_\Lambda\right) + \sum_{\gamma}{}'' z(\gamma) \left(\langle \chi(kl) \rangle_{\Lambda \setminus \gamma} - \langle \chi(kl) \rangle_\Lambda\right), \tag{50}$$

where the first sum is over defect clusters $\gamma$ which contain *both* $(ij)$ and $(kl)$, and the second over those containing $(ij)$ but not $(kl)$. The subscripts on the expectations indicate the lattice we are calculating for – the full lattice $\Lambda$, or with the cluster $\gamma$ removed, $\Lambda \setminus \gamma$.

If $K$ is small enough, the probabilities of $(kl) \in \gamma$ and of $|\gamma| > M$ asymptotically decay exponentially with the distance between $(kl)$ and $(ij)$ or $M$, respectively. Similarly, the difference $\langle \chi(kl) \rangle - \langle \chi(kl) \rangle_\Lambda$ decays exponentially with the distance of $(kl)$ from the boundary of $\Lambda$ (the former expectation is for an infinite lattice with no holes). If the correlation eqn. (50) is to exhibit power law decay at $K = K_c$, at least one of these three quantities must also show such a change in asymptotic behavior. It seems extremely unlikely that the effect of cutting out small pieces of the lattice has a qualitatively slower falloff than the cluster size. Accepting that, the transition at $K = K_c$, $J = 0$ is accompanied by a *percolation of defect lines*. The difference from ordinary bond percolation is that the frustration network is made of closed loops and thus is not allowed to have free ends.

This picture is easily related to the line tension. The correlation function of eqn. (49) and our manipulation of boundary conditions in section V both serve to to measure the ease with which a defect line can join two distant points. The line tension is the inverse of the correlation length for the frustration-percolation.

This percolation criterion must hold for $J > 0$ to be a useful description of the $I/T$ transition. There is little doubt that it is more robust against the perturbation of positive $J$ than either the Wilson loop or 't Hooft disorder operator. A crude estimate says that for $e^{-2K} < J$, the leading behavior of the large-$K$ defect expansion is the same as at $J = 0$. For the small-$K$ expansion in terms of plaquette-surfaces, the leading terms involve a tube running between the two plaquettes in question and the $J$ perturbation is negligible for $\tanh K > J^2$. As stated in section V, the line tension cannot vanish in phase *N*, since that would certainly destroy the long-range ordering. The case of the $\mathbb{Z}_2$ Higgs-gauge model [38], however, argues for some caution. It has only two phases – a free-charge phase at large $K$ and small $J$, and a confinement-screening phase everywhere else. The line tension, which vanishes in some region near $K = J = 0$, becomes positive at large $J$ without passage through a bulk phase transition.

### F. Small $K$

In section IV, we showed that for $K = 0$ our gauge theory becomes a spin-only model akin to the Lebwohl-Lasher model. Now we will use the methods introduced in the previous section to extend the elimination of the gauge variables to small non-zero $K$.

We expand the spin part of the Boltzmann factor as before, but without integrating over the spins, and use the pure gauge Hamiltonian to evaluate the expanded form term-by-term. Explicitly,

$$Z = \int [d\mathbf{S}]\, Z[\mathbf{S}], \tag{51}$$

with

$$Z[\mathbf{S}] = \prod_{\langle ij \rangle} \cosh(J\mathbf{S}_i \cdot \mathbf{S}_j) \times \left\langle \prod_{\langle ij \rangle} (1 + U_{ij} \tanh(J\mathbf{S}_i \cdot \mathbf{S}_j)) \right\rangle_G . \tag{52}$$

Equivalently, this can be written as

$$Z[\mathbf{S}] = Z[\mathbf{S}; K = 0] \sum_C Z_C(K) \prod_{\omega \in C} \rho(\omega, \mathbf{S}), \tag{53}$$

where the sum is over collections $C$ of closed graphs on the lattice, as in equation 26 for the large $K$ case, and $\rho(\omega, \mathbf{S})$ is defined in eqn. (25). The factor $Z_C(K)$ is the pure gauge theory partition function in the presence of source $C$ (eqn. 27). The gauge field induces a weak non-contact interaction between the spin polymers.

By use of eqn. (23), the Boltzmann factors associated with the pure gauge theory can also be rewritten as

$$e^{KU_P} = \cosh K (1 + U_P \tanh K), \tag{54}$$

where we use the shorthand $U_P = \prod_{(ij)} U_{ij}$ for the product of link variables around a plaquette $P$. Expanding the product gives us collections of plaquettes with factors of $\tanh K$. The expansion of the spin part is as before. Upon summing over $\pm 1$ for each link variable $U_{ij}$, any surviving term of an expansion must have an even number of occurrences of each link variable. Thus, our polymers will consist of edge-connected collections of plaquettes each carrying a factor $\tanh K$, and a factor $\tanh(J\mathbf{S}_i\mathbf{S}_j)$ for each boundary link $(ij)$, if any. In the presence of a Wilson loop, the loop and all the plaquettes from expansion of the gauge Hamiltonian or links from the spin Hamiltonian which overlap it on a link are to be considered as a single polymer, whose activity is zero unless all the factors of $U_{ij}$ are cancelled in pairs, and is otherwise evaluated as for the others.

The leading terms contributing to a Wilson loop in this case are (i) the one containing all the plaquettes on a surface bounded by the loop, which gives an area law decay in the case of $J = 0$, or (ii) the spin polymer which tracks along the Wilson loop, which gives a perimeter law and is dominant for large enough loops at $J > 0$. In either case, what determines the minimal polymer is the need to cancel the factors of $U$ along the loop. In the expansion of eqn. (53), the expectations are in the pure gauge Hamiltonian, so there is area law decay and the expansion is well under control. As a first approximation, we keep only single plaquette graphs. Exponentiating, the result is

$$H^{(1)}_{\text{eff}} = \sum_{\langle ij \rangle} \log[\cosh(J\mathbf{S}_i \cdot \mathbf{S}_j)] + J' \sum_{\square} \prod_{kl \in \square} \tanh(J\mathbf{S}_k \cdot \mathbf{S}_l). \tag{55}$$

Here, $J' = J^4 \tanh K$. Notice that the correction term in equation (55) is almost the same as the disclination counter equation (8), just without the sign function.

## VIII. OTHER PERTURBATIONS

The model we have been considering contains some *exact* symmetries which are only approximate for real nematics. These are (1) *local* head-to-tail symmetry, *i.e.* $\mathbf{S} \to -\mathbf{S}$, and (2) global space-spin rotation invariance, under which all the spins are subjected to the same rotation ( $\mathbf{S}(\mathbf{r}) \to \mathbf{R}\mathbf{S}(\mathbf{r})$). Local head-to-tail symmetry is lacking despite the fact that *global* head-to-tail symmetry is unbroken, *i.e.* that the heads and tails do not order. The absence of a globally broken symmetry does not imply that the Hamiltonian lacks terms breaking it locally – only that they are too weak to induce long range order (vector ordering in this case).

The absence of global spin-rotation invariance is exhibited explicitly in the Frank free energy (eqn. 1) for generic $K_i$ (it *is* invariant at the special point $K_1 = K_2 = K_3$). Since the spins of our model are related to the physical orientation of extended bodies, it is not surprising that the only exact symmetry is that under simultaneous and identical rotations in spin- and real-space.

We show here that small perturbations of the Hamiltonian which break these symmetries do not change our results, so that we can feel safe applying them to real-world nematics.

### A. Lack of Local Inversion Symmetry

First, we consider the local head-to-tail symmetry. We need an extra variable $\alpha_i = \pm 1$ for each site which keeps track of whether $\mathbf{S}_i$ is oriented with the local vector. It has identical gauge transformation properties to that of $\mathbf{S}_i$ itself, so that the product $\alpha_i \mathbf{S}_i$ is the "real" local vector (and gauge invariant as it must be). A direct Heisenberg interaction is added to the Hamiltonian of eqn. (11) by the term $g \sum_{\langle ij \rangle} \alpha_i \alpha_j \mathbf{S}_i \cdot \mathbf{S}_j$, for some small $g$ (compare the "Mattis spin-glass"), where the $\alpha$'s are to be summed over in the partition function. As far as the spins $\mathbf{S}$ are concerned, this is equivalent to adding a second independent gauge field with $K = \infty$, thus the Heisenberg character of the transition is clearly preserved. It is also easy to see that the interaction does not produce a true vector ordering once the nematic phase is entered. The tendency toward complete vector ordering is only strengthened if all the spins are forced to orient along a common axis, yielding an ordinary Ising model in the variables $\alpha_i \mathbf{S}_i$. But the coupling $g$ is weak, so there will be no ordering.

### B. Space-spin Coupling

Near the $N/T$transition we have seen that our model can be mapped onto an effective Heisenberg model. In the critical regime, a familiar transformation [1] converts this fixed spin model into an $n = 3$ "soft-spin" $|\mathbf{S}|^4$ Landau theory with Hamiltonian density

$$\mathcal{H} = \frac{1}{2}|\nabla \mathbf{S}|^2 + \frac{1}{2} r |\mathbf{S}|^2 + u|\mathbf{S}|^4, \tag{56}$$

where $\mathbf{S}$ is now a three-vector with unconstrained length. To incorporate space-spin couplings, we introduce terms which mix up the spatial component indices and the spin components. The only such terms with any chance of altering the critical behavior have as few gradients and powers of $\mathbf{S}$ as possible. Indeed any term involving more than two powers of each will be strongly irrelevant. The only potentially relevant term is therefore $|\nabla \cdot \mathbf{S}|^2$. This perturbation was analyzed [39] in the heyday of RG in an $\epsilon = 4 - d$ expansion, and found to be irrelevant, with renormalization group eigenvalue $\lambda = -\epsilon^2/108 + \mathcal{O}(\epsilon^3)$. Thus, there is nothing to fear from space-spin couplings, either: the Heisenberg universality class of the transition survives. (This issue is discussed further in Paper II.)

## IX. MONTE CARLO RESULTS

### A. General Methods

We have employed Monte Carlo simulation to investigate the lattice gauge model of eqn. (11). The vast majority of runs were for three-component spins, though we have also investigated four-component spins. The simulations were implemented on a Sun SparcStation with a standard Metropolis algorithm on lattices of size up to 16×16×16 sites. Most of the runs employed periodic boundary conditions for all variables (spin and gauge) to eliminate boundary effects. A combination of free and fixed boundary conditions, however, was necessary to measure the line tension (see section IX C and V).

Instead of allowing the spin variables to sample the entire unit sphere, we used a discrete set of allowed values. This simplifies and speeds the simulations, since Boltzmann factors can be stored in a lookup table, and it is also somewhat easier to select candidate Monte Carlo moves. For the three-dimensional order parameter space, we used the most symmetric set of allowed vectors, namely the 30 vectors pointing to the centers of the edges of an icosahedron. Such a discretization represents an anisotropy of the single-spin weight which breaks the original $O(3)$ rotation invariance to its largest discrete subgroup, the icosahedral group $Y$.

In order for this anisotropy to be irrelevant at the Heisenberg critical point (in the renormalization group sense), the allowed spin orientations must to cover the sphere sufficiently uniformly. The icosahedral edge vectors easily pass [40] this test. Use of a coarser discretization which is still irrelevant presents two problems. First, it will be necessary to get closer to the transition before crossing over to the fully-symmetric critical behavior. Secondly, the discretization introduces a spurious freezing transition, due to the presence of a spin wave gap. It is desirable to push this artifact deep into the nematically ordered phase. The same effect also results in a small shift of the numerical value of the critical coupling.

Among the measurables which were extracted from the simulations were

1. the average plaquette value $\overline{P} \equiv \langle U_{ij} U_{jk} U_{kl} U_{li} \rangle$,
2. the gauge-invariant nearest-neighbor correlation $C \equiv \langle U_{ij} \mathbf{S}_i \cdot \mathbf{S}_j \rangle$,
3. average Wilson loop of several sizes, and
4. a scalar measure of the strength of nematic ordering given by

$$\begin{aligned} q^2 &\equiv \frac{N}{N-1} \left\langle \frac{3}{2} \mathrm{Tr}\ \mathrm{Q}^2 - \frac{1}{\mathrm{N}} \right\rangle \\ &= \frac{1}{N(N-1)} \frac{3}{2} \left\langle \sum_{i \neq j} \mathrm{Tr}\ \mathrm{Q(i)Q(j)} \right\rangle, \end{aligned} \tag{57}$$

where $Q_{\alpha\beta} \equiv (1/N) \sum_i Q(i)_{\alpha\beta}$ is the nematic order parameter, and $N$ is the number of sites in the lattice.

Tr $Q^2$ is invariant under rotation and measures the degree of alignment of the individual spin axes for a uniaxially ordered phase. The normalization of eq. (57) is chosen so that $q^2$ vanishes in the disordered phase and is equal to unity in the completely ordered state at zero temperature. Terms with $i = j$ are excluded from the definition of $q^2$; they would make a constant contribution subleading in $1/N$. Leaving them out ensures $q^2 = 0$ in a fully disordered state. Normalization is such that $q^2 = 1$ in a completely ordered state.

We studied the equilibrium values of these quantities by stepping along lines in the $J - K$ phase diagram at a variety of orientations, allowing the system to reach equilibrium after each incremental change, and then making measurements. The "temporal" development of these quantities for values of the couplings close to the transitions were examined by eye to determine the equilibration time. Thermalization typically required 700-2,000 update attempts per degree of freedom (link or site) with each coupling step. Typically 6,000 to 15,000 independent measurements were made at each coupling value, spaced by two or three update sweeps through the lattice to give reasonable statistical independence. This was sufficient to extract equilibrium averages of the quantities (1-4). To obtain accurate results on fluctuations about equilibrium values (*e.g.* the specific heat), as well as to examine finite-size scaling, 30,000-60,000 measurements were required at each coupling step.

Our data confirm the existence of the three phases which were predicted analytically in previous sections. They also show that the transition between phases $N$ and $I$ is first-order, and the transitions between $N$ and $T$ and between $I$ and $T$ are continuous. As is well known, however, it can be difficult to definitively establish the order of a transition *via* Monte Carlo. We have used finite-size scaling for the continuous transitions and shown phase coexistence at the first order $N/I$ boundary.

The distribution of Tr $Q^2$ for values of $J$ near the $N/I$ transition shows it to be first-order at $K = 0$. The double-peaked distribution shown in figure 7 demonstrates the coexistence of ordered and disordered phases. Clear evidence of a discontinuity in the order parameter expectation value $q^2$ is also seen, a feature which becomes sharper with increasing lattice size.

Figure 8a displays the development of the order parameter $q^2$ near the ordering transition close to the point where it meets the Ising transition line. The data strongly suggests that the nematic ordering transition is first order to the left of that point and continuous to its right. The mean plaquette value and specific heat also show qualitatively different behavior from one side of that point to the other. The difference is discernible even between $K = 0.73$ (on the first order side) and $K = 0.77$ (continuous). On the first-order side, the specific heat peak sharpens rapidly as the lattice size is increased, compared to a much more gradual sharpening for larger $K$ at which the transition is continuous. This behavior of the specific heat is shown in figure 8b (see also the next

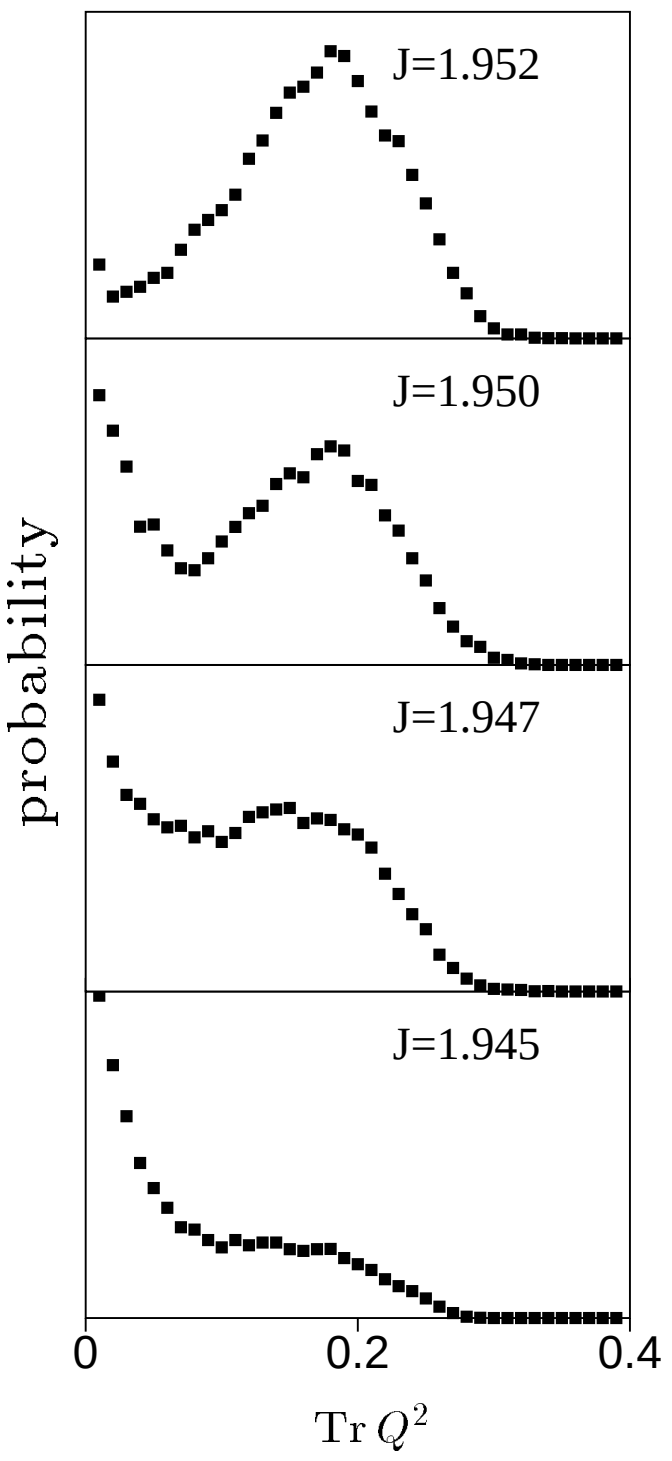

FIG. 7. Probability distributions for order parameter at $K = 0$ in a $12^3$ lattice for several values of $J$ near the first-order transition. As $J$ increases, the development of a bump corresponding to the ordered state and consequent diminishing of the disordered-state bump at zero order parameter is evident. In the thermodynamic limit, at any given $J$, only one of the bumps is expected to survive. The distributions were constructed by making histograms of a Monte Carlo "time" series.

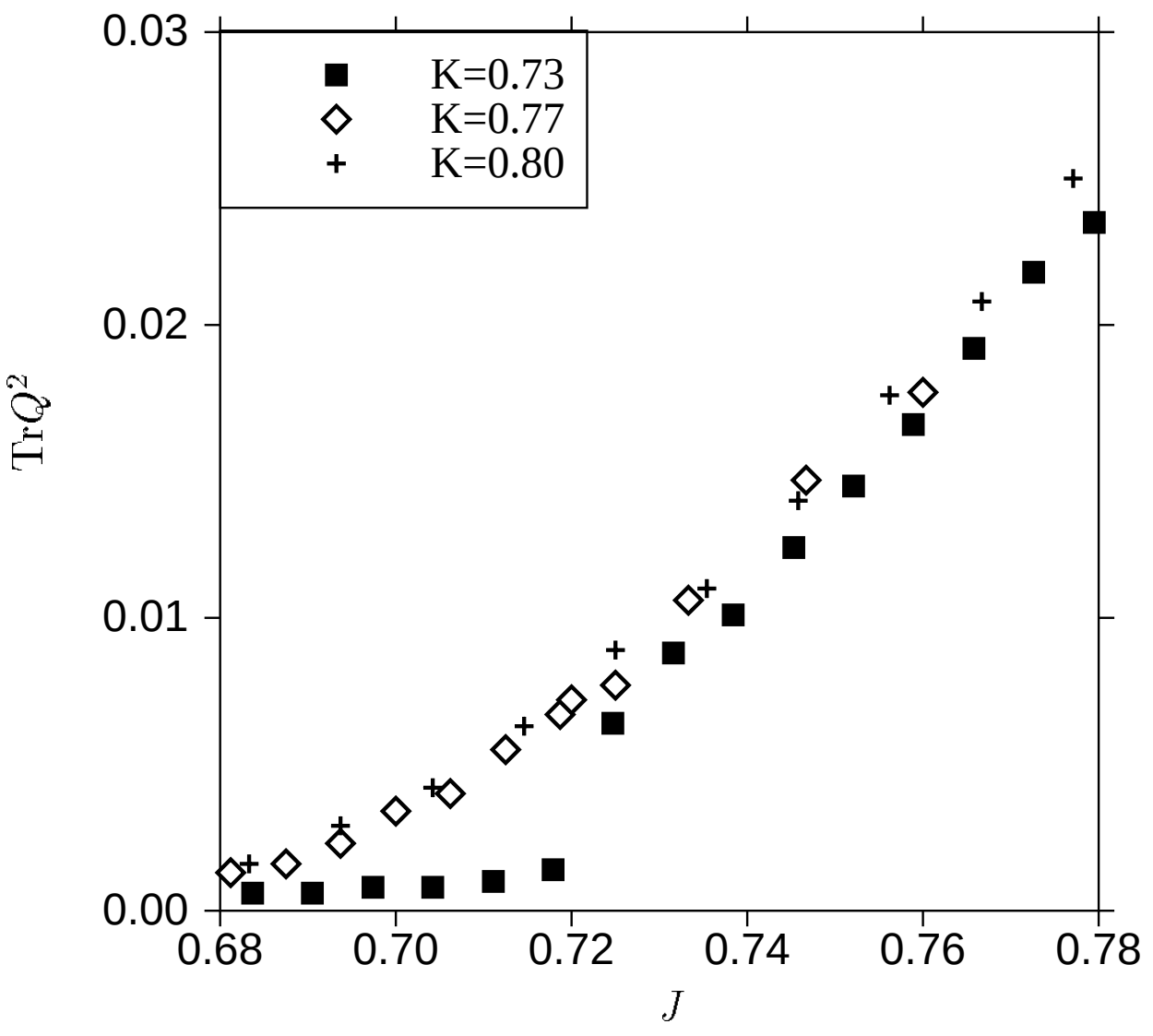

FIG. 8. Development of the nematic order parameter Tr $Q^2$ at the transition, and near the crossover point from first-order to continuous. All data are for a $12^3$ lattice. The apparent sharpness of the jump for $K = .73$ is essentially accidental; neighboring data points straddle the region where Tr $Q^2$ begins to rise from zero.

section and figure 10).

### B. Finite-Size Scaling

Finite-size scaling analysis is a well-developed tool for the determination of critical exponents and orders of transitions (for reviews, see references [41] and [42]). The finite-size scaling *ansatz* for critical points asserts that in finite geometries characterized by a linear dimension $L$, scaling forms depend on a scaling variable $L/\xi$, in addition to those already present at infinite system size. For a magnetic system, the critical coupling can be determined from the ratio

$$f_n(L,t) \equiv \frac{\langle m^{2n}\rangle_L}{\langle m^2\rangle_L^n} \approx f_n(Lt^\nu). \qquad (58)$$

Here, $m$ is the spontaneous magnetization, and $n$ is a small integer. The indicated functional dependence follows since both numerator and denominator have the same scaling dimension. Thus, at the critical temperature, this quantity is *independent of* $L$. Furthermore,

$$\lim_{t\to 0} \frac{d}{dt} f(Lt^\nu) \propto L^{1/\nu}, \qquad (59)$$

as can be seen by noting that this derivative must be finite and non-zero as $t \to 0$ at fixed $L$. Since $df(Lt^\nu)/dt = \nu L t^{\nu-1} f'(Lt^\nu)$, we must have $f'(x) \propto x^{1/\nu - 1}$ as $x \to 0$, leading immediatedly to equation (59). This relation allows determination of $\nu$.

At $K = \infty$, our model has the same partition function as the Heisenberg ferromagnet, and we expect to have the same exponents at the transition for $K$ down to the multicritical value. Since the magnetization is not a gauge-invariant quantity, we instead consider

$$g(L,t) \equiv \frac{\langle (\mathrm{Tr}\ \mathrm{Q}^2)^2\rangle_\mathrm{L}}{\langle \mathrm{Tr}\ \mathrm{Q}^2\rangle_\mathrm{L}^2} = g(Lt^\nu), \qquad (60)$$

analogous to $f_4(L,t)$ (equation 58, with $h = 4$). We use Tr $Q^2$ here, rather than $q^2$ (equation 57), because the analogous magnetic quantity, $m^2 = (N^{-1}\sum S_i)^2$, is nonzero at infinite temperature. Indeed, the presence of terms $|S_i|^2 = 1$ in Tr $Q^2$, but not in $q^2$ (recall the discussion at the beginning of section IX A) result in $m^2 \sim 1/N$ at infinite temperature. Splitting off the $i = j$ terms, each of which is $|S|^2 - 1/3 = 2/3$,

$$\begin{aligned}\langle \mathrm{Tr}\ \mathrm{Q}^2\rangle_\mathrm{L} &= N^{-2}\sum_{i\neq j}\langle (S_i\cdot S_j)^2 - 1/3\rangle_L + 1/N \\ &= \langle q^2\rangle_L + N^{-1}(1-\langle q^2\rangle_L), \end{aligned} \qquad (61)$$

which tends to $1/N = 1/L^3$ as $J \to 0$. The numerical data for this function for $L = 8$, 12, and 16, show intersections indicating a value of the critical coupling between 0.690 and 0.695, in agreement with the best value to date for the three-dimensional Heisenberg model [20], $J_c(K=0) = 0.693...$ The shift of the critical coupling due to icosahedral anisotropy is evidently small. At $K = 0.78$, the critical coupling is a bit larger: $J_c(K = 0.78) = 0.70$.

For a magnetic system, the finite size scaling of the magnetization is also quite useful:

$$m = |t|^\beta f(L/\xi) = L^{\beta/\nu} f(Lt^\nu). \qquad (62)$$

According to eqn. (62), the ratio $\beta/\nu$ is accessible by measuring the magnetization, which can be determined accurately. Our order parameter is related to the magnetization of an equivalent Heisenberg model, though not quite the same. For a finite lattice at values of $J$ not too much below $J_c$, the approximation [43]

$$\langle (S_i\cdot S_j)^2\rangle_L = \langle (S_i^\alpha S_i^\beta)(S_j^\alpha S_j^\beta)\rangle_L \approx \langle S_i^\alpha S_i^\beta\rangle_L \langle S_j^\alpha S_j^\beta\rangle_L, \qquad (63)$$

should be valid for $i$ and $j$ well-separated. The last expression in (63) is proportional to $m^4$. Notice that as $J \to 0$, $m^4 \sim 1/N^2$, but $\langle \mathrm{Tr}\ \mathrm{Q}^2\rangle \sim 1/\mathrm{N}$, due to the $i = j$ terms in eqn. (61). Thus, $\langle q^2\rangle_L$ (which lacks these extra terms) will behave much more like $m^4$ than $\langle trQ^2\rangle_L$ does. We conclude that

$$\langle q^2\rangle_L \approx L^{4\beta/\nu} f(Lt^\nu). \qquad (64)$$

At the critical coupling (determined from eqn. 58), eqn. (64) can be used to determine $\beta/\nu$. Combining with $\nu$ from eqn. (59) gives $\beta$. Repeating this analysis at $J = 0.693$, $K = \infty$, we found $\nu = 0.71 \pm 0.04$, $\beta = 0.39 \pm 0.03$, to be compared with the accepted Heisenberg values of 0.703 and 0.38 respectively. At $K = 0.78$, we find $\nu = 0.72 \pm 0.05$ and $\beta = 0.4 \pm 0.04$, the same to within error. The scaling in equation (64) can also be checked away from the critical coupling, by plotting $q^2 L^{-4\beta/\nu}$ versus $L^{1/\nu}(J - J_c)$. Tn data collapse is best for the Heisenberg values. One expects that effects of a finite gauge coupling should be observable at $K = 0.78$, if they are relevant, which they do not appear to be. Scaling plots of the order parameter at $K = 5$ and $K = 0.78$ are shown in figure 9.

A finite-size scaling analysis was also carried out for the specific heat at $K = 0.70$. The specific heat curves are shown in figure 10. Finite-size scaling near a first-order transition is quite different from that near a critical point [44]. At the discontinuity fixed point governing a first-order transition, the only relevant exponent [45] is the spatial dimensionality $d$ of the system. One therefore expects to see no exponent other than $d$ in the leading-order finite size scaling formulas for a first-order transition, in contrast to those for a continuous transition which contain the familiar anomalous exponents. The rounding of the delta-function in the specific heat should thus result in a peak height linear in the volume of the system, *i.e.*

$$C(L,T) = C_{\text{smooth}}(T) + L^d f\left(\frac{T - T_c}{L^d}\right). \qquad (65)$$

The same conclusion follows [46] by assuming that configurations of the finite system occur in the partition function with a weight which is the average of those appropriate for the two coexisting phases. The data for $K = 0.70$ appearing in figure 10a are consistent with such a scaling. The data for $K = 0.78$ (figure 10b), however, do not appear consistent with linearity, indicating that the transition is *not* first order at that value of $K$.

### C. Line Tension

We also measured the defect line tension in our Monte Carlo simulations. The boundary conditions (a) and (b) of section V are relatively straightforward to implement in the lattice gauge theory of eqn. (11). In two lattice directions ($\hat{\mathbf{x}}$ and $\hat{\mathbf{y}}$, say), the boundary conditions are open for the spins. The product of the link variables circling once around the periphery in the XY plane is required to be either +1 (boundary condition a) or −1 (boundary condition b). In the remaining lattice directions (just $\hat{\mathbf{z}}$ in three dimensions), the boundary conditions are chosen open with the restriction that either no boundary plaquettes are frustrated (a), or only the central plaquettes on the top and bottom faces are frustrated (b). The free energy difference per unit length in the z-direction as $L_x$, $L_y \to \infty$ is interpreted as a disclination free energy.

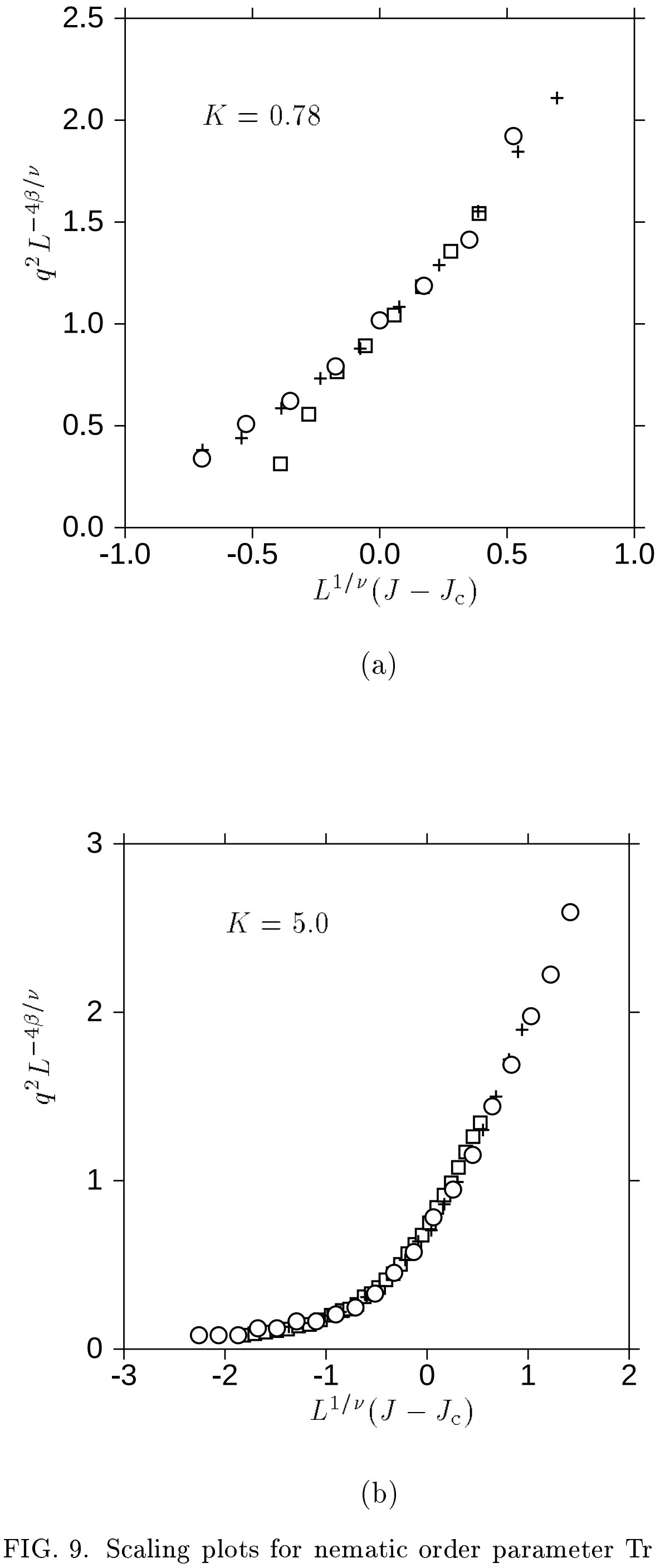

FIG. 9. Scaling plots for nematic order parameter Tr $Q^2$. (a) $K = 0.78$, with $\nu = 0.7$, $\beta = 0.4$. (b) $K = 5$ (effectively infinite), with $\nu = 0.7$, $\beta = 0.38$. The lattice sizes are: $L = 8$ (squares), $L = 12$ (crosses), and $L = 16$ (circles).

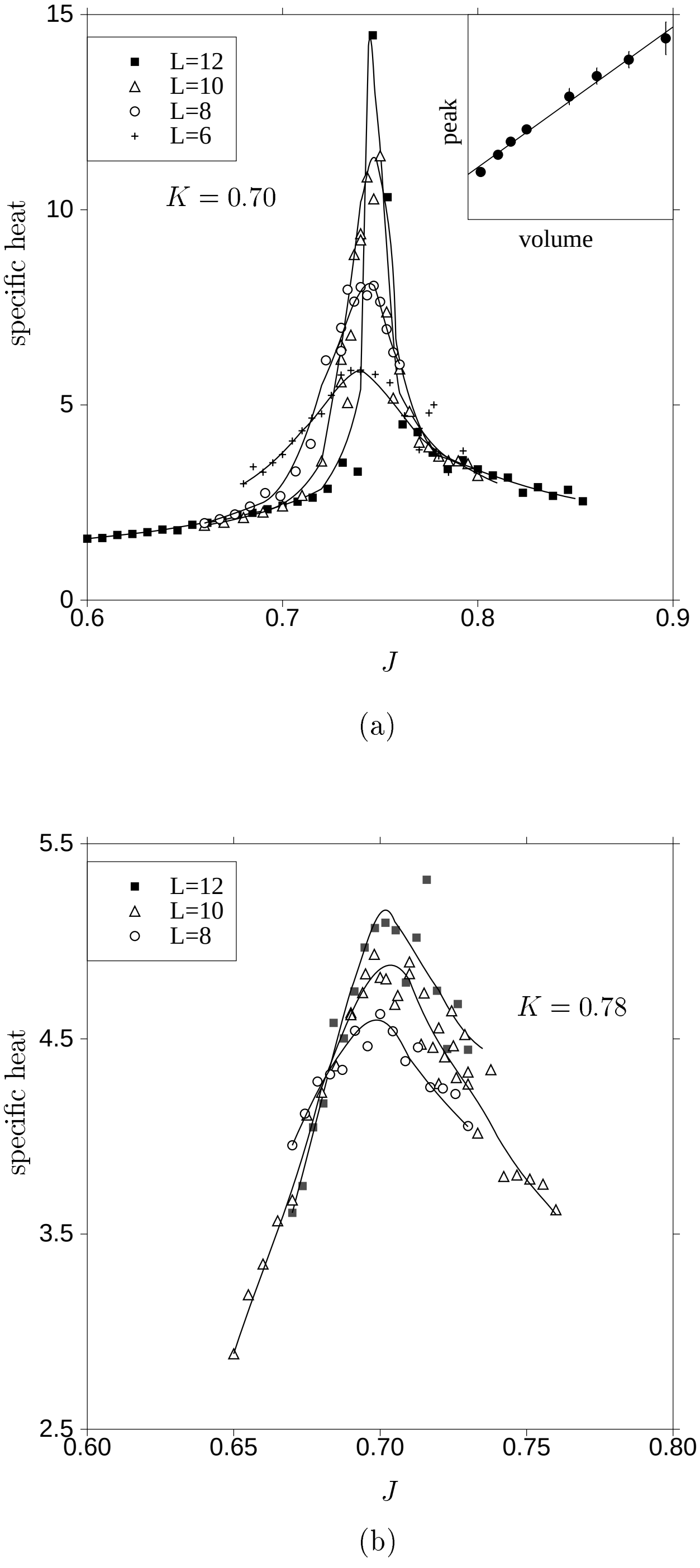

FIG. 10. Evolution of specific heat peaks with lattice size in (a) first-order ($K = 0.70$) and (b) continuous ($K = 0.78$) transition regions. In the first order region, the peak sharpens very rapidly with increasing lattice size. The linear scaling of peak height with lattice volume is shown in the inset. The continuous curves are guides to the eye only.

Essentially this construction was suggested as a diagnostic for quark confinement by Mack and Meyer [36] in the context of the the $\mathbb{Z}_2$ Higgs-gauge model.

The general strategy is to force a defect to traverse the system in the $z$-direction, going through the centers of the faces at $z = 0$ and $z = L_z$, but to forbid defects from passing through the boundary anywhere else. This means fixing the gauge field strength everywhere on the boundary. The simplest means of doing this is by actually partially fixing the gauge, freezing the link variables on the boundary into an appropriate configuration and not altering them during the simulation. It is also possible to use periodic boundary conditions in the $z$-direction and still force a defect to run through, so that a Wilson loop going around the boundary and circling the $z$-axis is $-1$. The defect is attracted to the boundary by image forces, however, where it produces the least disturbance, which needs to be avoided. Since the spins do not need to be constrained in any way, free boundary conditions are used for them.

The line tension can be studied in such a setup by varying $L_z$ at fixed $L_x$ and $L_y$. As discussed earlier, in the infinite volume limit the line tension is expected to vanish at $K_c$ with the correlation length exponent $\nu_I$ of the Ising model. Finite size effects make interpretation of the data difficult. The problem is also exacerbated by working on a lattice with boundaries. We measured the line tension at $J = 0.5$ as a function of $K$. This value of $J$ is chosen to be in the nematically disordered region, yet with large enough nematic coupling $J$ that its effects ought to be observable. The free energy is determined by integrating the plaquette expectation value with respect to $K$. In principle we should integrate from the corner $J = K = 0$, but we actually start from $K, J = (0, 0.5)$, since within statistical error $\partial(\delta F)/\partial J = 0$ at that point, hence *a fortiori* for all smaller $J$. The evolution of the line tension with $K$ at $J = 0.5$ is shown in figure 11. There is a kink near the value of $K$ at which the specific heat peak occurs. This kink sharpens and moves downward with increasing $L_z$, indicating that the tension vanishes above $K_c$ in the infinite volume limit. The derivative of the line tension with respect to $J$ across the $N/I$ transition at $K = 0.5$ is shown in figure 12. As the lattice size increases, the peak is sharpening into a delta-function reflecting the strictly positive value of the nematic order at the transition. Apart from observing the qualitative trend with system size, we have done no quantitative finite-size scaling analysis. Multiple length scales resulting from non-periodic boundary conditions makes such an analysis impractical.

### D. Defect Line Statistics

We have also examined the configurations of defect lines in our simulations. Figure 13 depicts typical defect networks for several representative points in the phase diagram. Upon passing into the totally disordered confinement phase $I$ from the free charge phase $T$, a dramatic increase in the number of defect loops is evident. The increase in total length is not quite so obvious in the $N/I$ transition because even in the nematic phase there are already many small loops near the transition. When the lattice becomes crowded with defective plaquettes, it is impossible to unambiguously pick out individual *loops* of defect. The loops coalesce into defect clusters because of their intersections. We have compiled some statistics for these clusters. Figure 14 shows the distributions of total length for the defect clusters in typical configurations as $J$ is varied at $K = 0$. The $N/I$ transition is characterized not only by a sudden increase in the total length of defect, but also by the appearance of very long defect loops. The mean square separation of dual lattice sites on the clusters was also studied; in a toroidal lattice this is the quantity most nearly equivalent to radius of gyration. For clusters with radii smaller than the length of the lattice, the mean-square separation scales approximately as net length, as appropriate for a random walk. Any self-avoidance of the defect clusters is apparently weak.

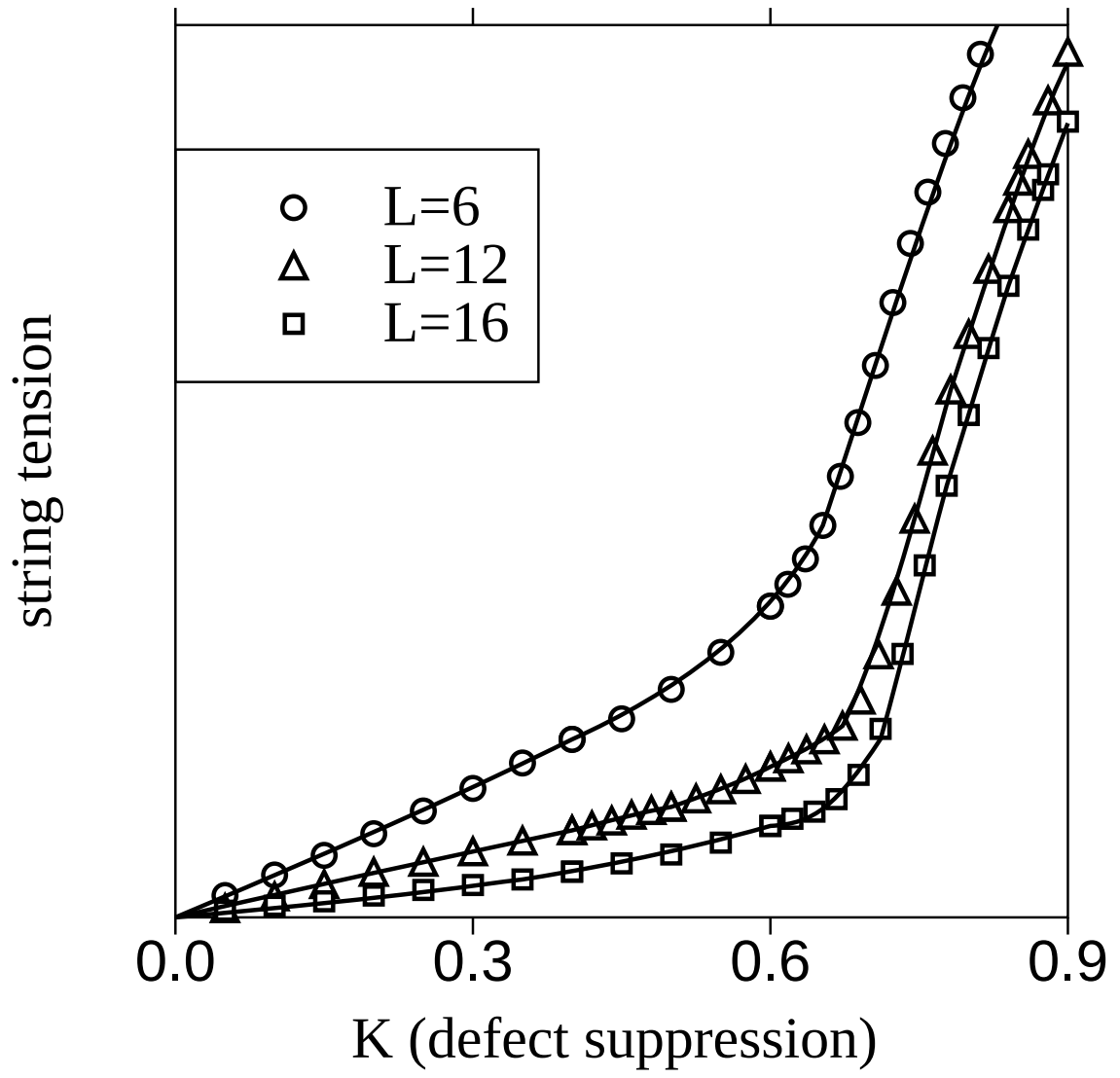

FIG. 11. The free energy difference $\delta F/L$ between a pair of $12 \times 12 \times L$ rectangular prisms with and without an externally imposed defect, respectively, as a function of $K$ at $J = 0.5$. Lattice sizes are $L$=6 (circles), 12 (triangles), and 16 (squares). Below $K \simeq 0.7$, the line tension is zero in the thermodynamic limit.

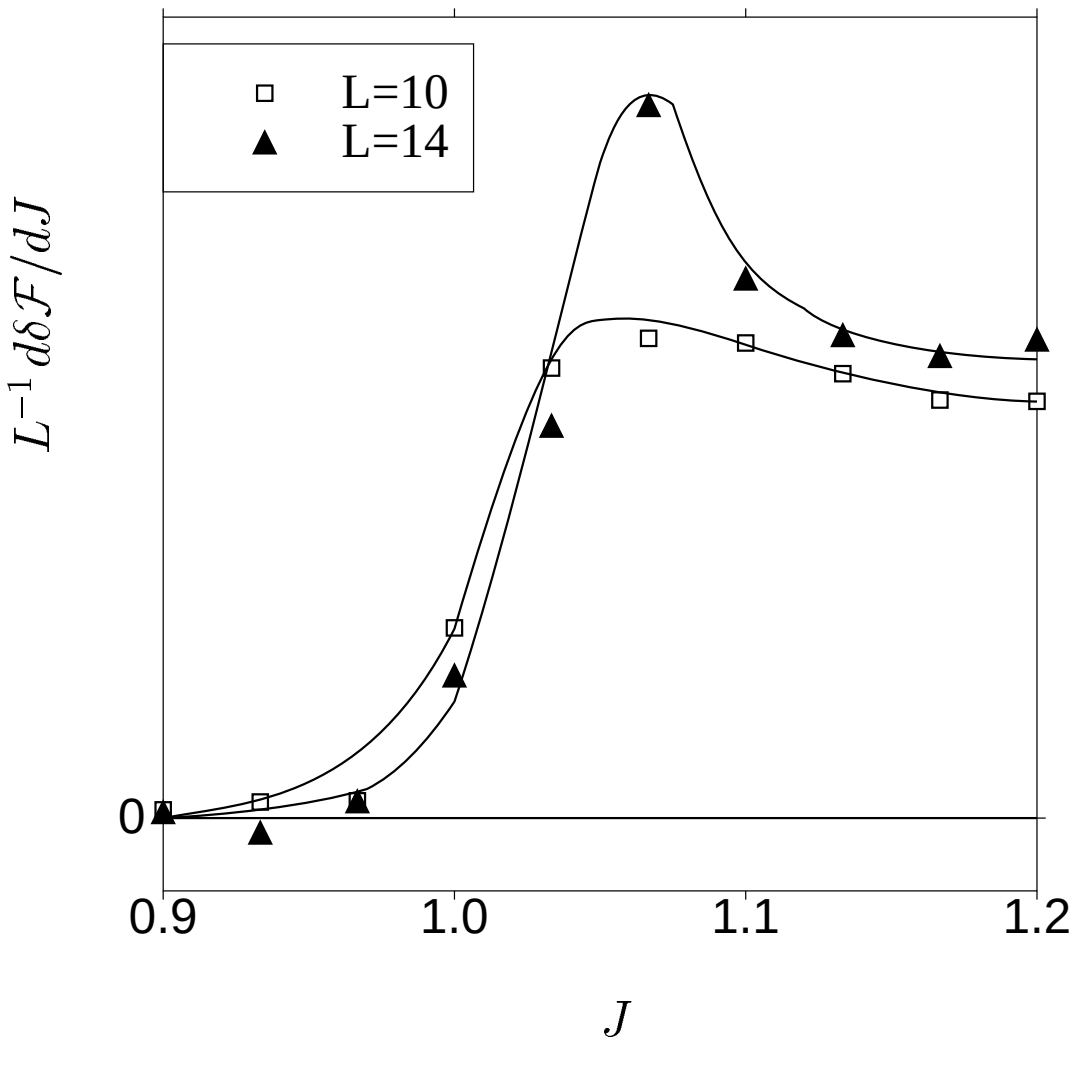

FIG. 12. Derivative of the line tension with respect to $J$ across the $N/I$ transition at $K = 0.5$. The squares are for lattice size $10^3$ and the triangles for $14^3$.

### E. Higher (Spin)-Dimensional Models

As indicated at the beginning of this section, we have also carried out simulations for the system with four-component spins situated at the lattice sites. The phase diagram for this case is qualitatively similar, with the ordering transition shifted toward larger $J$. Note that for a four-component nematic point defects are excluded by topological considerations ($\pi_3(\mathbb{RP}^3) = 0$), so that disclinations are the only allowed type of defect. Since there is no qualitative distinction, we have not pursued a full analysis.

## X. CONCLUSION

The ordered states of ferromagnetic and nematic media are strikingly similar, yet the experimentally observed ordering transitions of the two are quite different. Nematic transitions are observed to be weakly first-order and ferromagnetic ones are continuous.

In this paper, we have shown that the origin of this difference lies in the topological differences between the appropriate order parameter spaces; specifically, the presence of disclination lines in the nematic. Furthermore, we have shown that this effect can be completely suppressed by finite, short-ranged interactions, thereby making the disordering transitions of the two systems identical. In the process, we have discovered a new phase of nematogenic materials – the phase entered when the nematic disorders via a ferromagnet-like, continuous Heisenberg transition. This phase exhibits topological order which is destroyed only at a second, distinct transition in the Ising universality class. The full scenario is summarized in the phase diagram (figure 3). Experimentally observable critical behavior at the two continuous phase transitions in that figure are worked out in detail in Paper II. The problem of formulating a Landau-Ginzburg description of our theory and its relation to the usual Landau-Ginzburg theory [2] of the nematic will be addressed in a future publication.

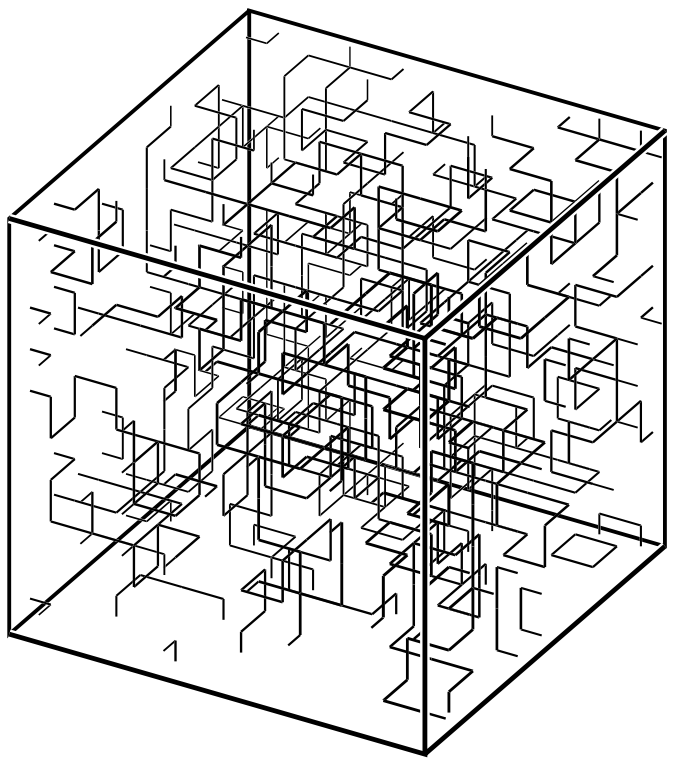

(a)

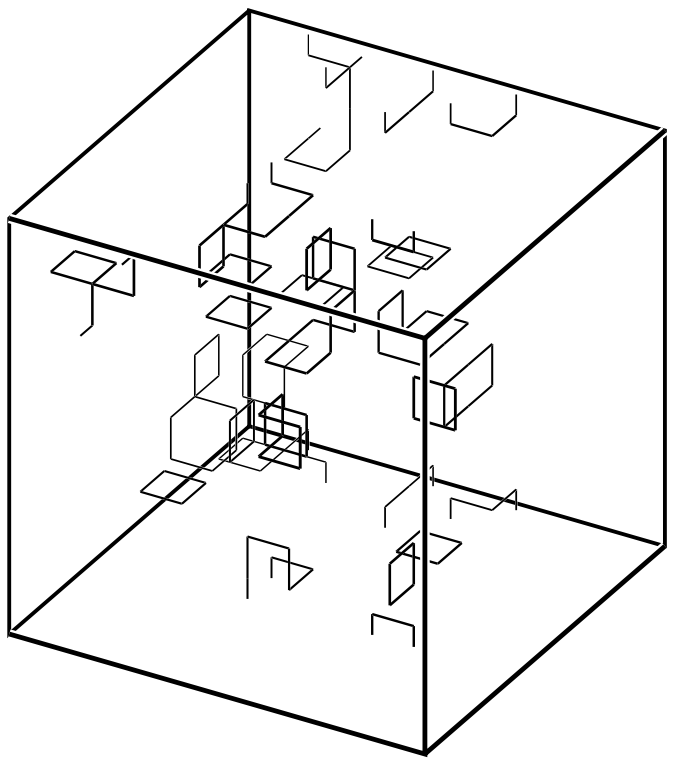

(b)

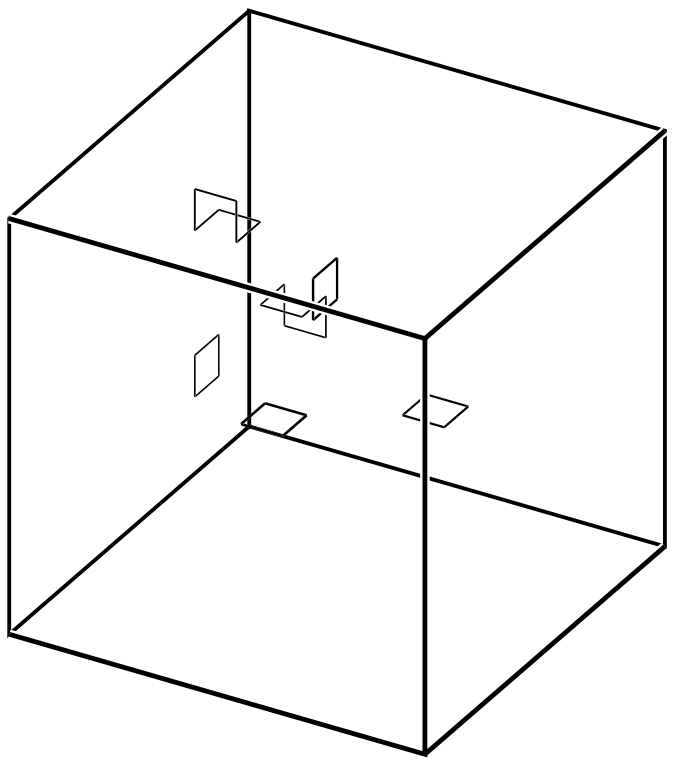

(c)

FIG. 13. Snapshot pictures of typical defect configurations. (a) in the nematic ($N$) phase at ($J = 1.3, K = 0.4$), (b) in the isotropic ($I$) phase at ($J = 1.1, K = 0.4$), and (c) in the isotropic $T$ phase, at ($J = 0.4, K = 0.85$). Periodicity of the boundary conditions is revealed by some of the defects in (a) and (b).

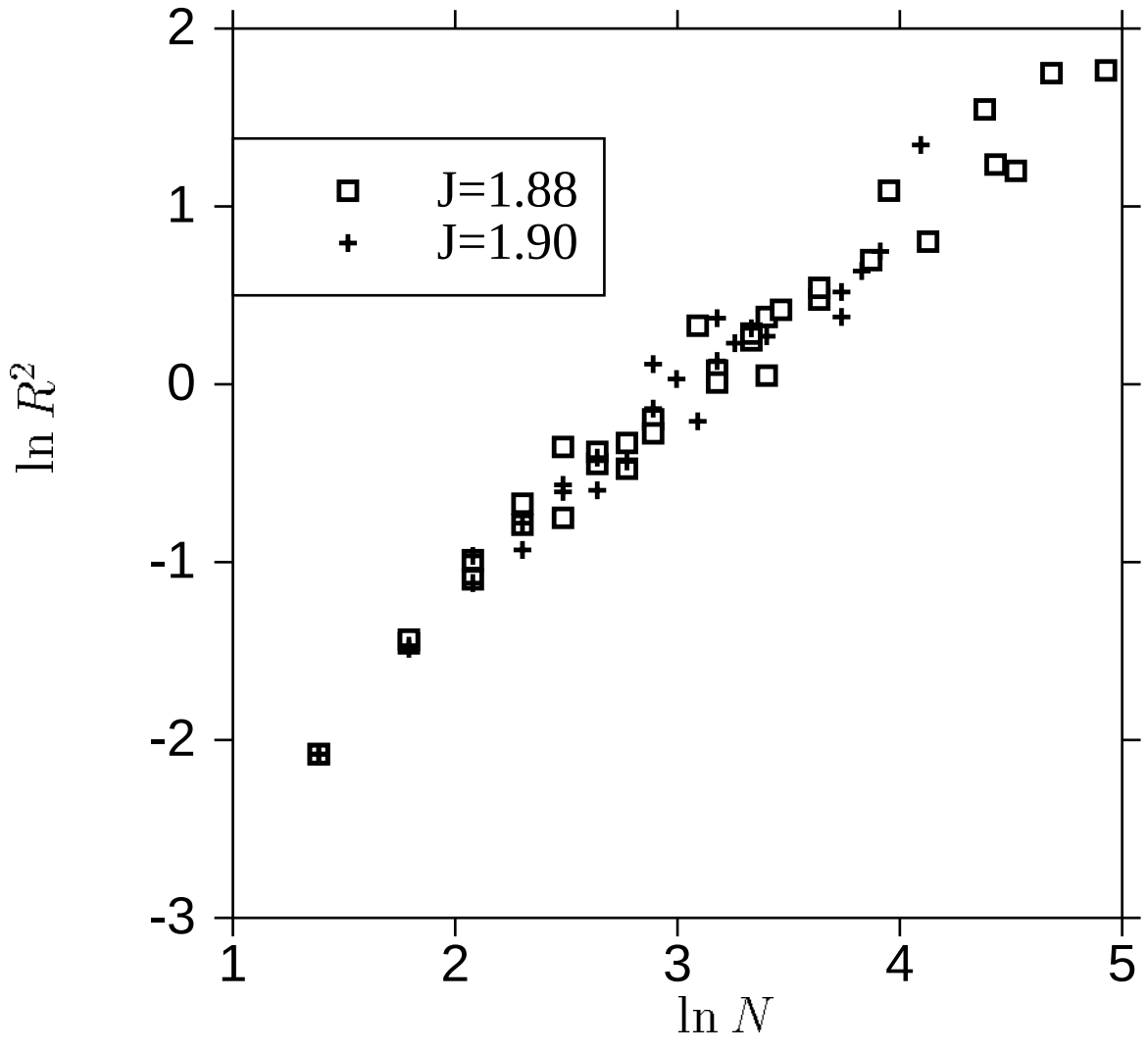

(a)

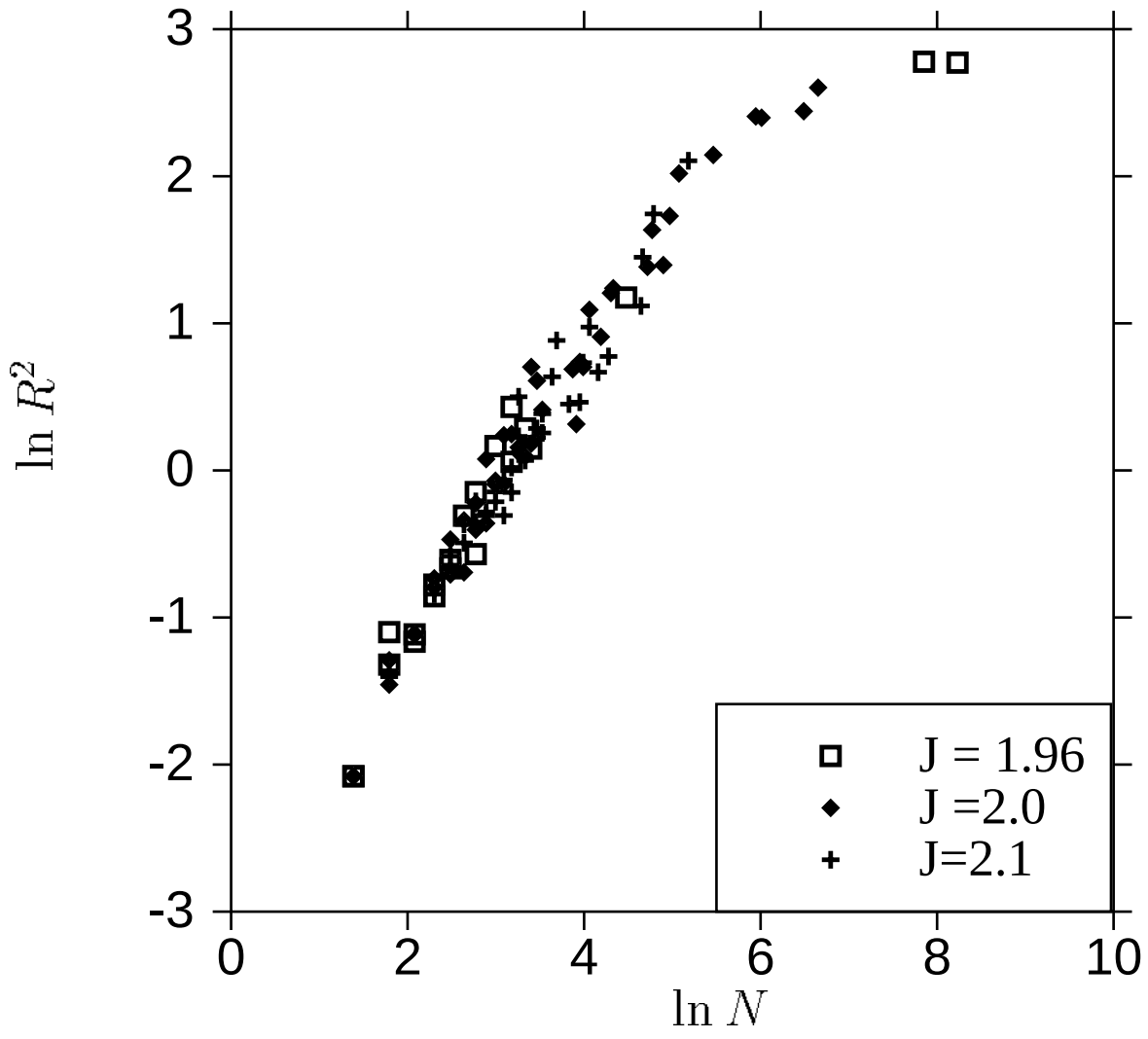

(b)

FIG. 14. Scaling of the radius of gyration ($R^2$) of defect clusters with total number of links ($N$) near the first-order transition at $K = 0$. As $J$ is lowered toward the transition value, there is a proliferation of defect loops (both gauge defects and disclinations) of all lengths. The points fall close to the line corresponding to a random walk for all values of $J$, except for some very large clusters, which have wrapped completely around the toroidal lattice. The two plots correspond to different definitions of defect: (a) is computed from eqn. (8) and (b) from the second term of the Hamiltonian in eqn. (11). The data (a) was taken with only 20 discrete values of the spin because it is necessary that there are no spin values with $S_i \cdot S_j = 0$; consequently the value of $J$ at the transition is somewhat smaller for this case.

## ACKNOWLEDGMENTS

JT thanks D. Roux for many discussions of his experiment, J. Prost for pointing out the possible connection of those experiments to this work, and for hospitality while various portions of this work were underway, the Aspen Center for Physics, the CNRS Paul Pascal (Bordeaux, France), the Isaac Newton Institute for Mathematical Sciences (University of Cambridge, UK) and the Institute for Theoretical Physics of U.C. Santa Barbara (and their NSF Grant PHY89-04035). PL acknowledges IBM and the U.S. Dept. of Education for fellowship support. DSR thanks V. Luby for interesting conversations and acknowledges grant support from NSF PYI 91-57414 and the Sloan Foundation.